\pdfoutput=1
\RequirePackage{ifpdf}
\ifpdf 
\documentclass[pdftex]{sigma}
\else
\documentclass{sigma}
\fi

\usepackage{bbold}
\newcommand{\Case}[2]{{\textstyle \frac{#1}{#2}}}
\newcommand{\lP}{\ell_{\mathrm P}}
\newcommand{\sgn}{\mathrm{sgn}(e)}

\def\k{\mathbf{k}}
\def\x{\mathbf{x}}
\def\y{\mathbf{y}}

\numberwithin{equation}{section}

\begin{document}

\allowdisplaybreaks

\renewcommand{\thefootnote}{$\star$}

\renewcommand{\PaperNumber}{010}

\FirstPageHeading

\ShortArticleName{Matter in Loop Quantum Gravity}

\ArticleName{Matter in Loop Quantum Gravity\footnote{This
paper is a contribution to the Special Issue ``Loop Quantum Gravity and Cosmology''. The full collection is available at \href{http://www.emis.de/journals/SIGMA/LQGC.html}{http://www.emis.de/journals/SIGMA/LQGC.html}}}

\Author{Ghanashyam DATE~$^\dag$ and Golam Mortuza HOSSAIN~$^\ddag$}

\AuthorNameForHeading{G.~Date and G.M.~Hossain}

\Address{$^\dag$~The Institute of Mathematical Sciences, CIT Campus, Chennai, 600 113, India}
\EmailD{\href{mailto:shyam@imsc.res.in}{shyam@imsc.res.in}}

\Address{$^\ddag$~Department of Physical Sciences,
Indian Institute of Science Education and Research - Kolkata,\\
\hphantom{$^\ddag$}~Mohanpur Campus, Nadia - 741 252, WB, India}

\EmailD{\href{mailto:ghossain@iiserkol.ac.in}{ghossain@iiserkol.ac.in}}

\ArticleDates{Received October 16, 2011, in f\/inal form February 27, 2012; Published online March 09, 2012}

\Abstract{Loop quantum gravity, a non-perturbative and manifestly background free,
quantum theory of gravity implies that at the kinematical level the
spatial geometry is discrete in a specif\/ic sense. The spirit of
background independence also requires a non-standard quantum
representation of matter. While loop quantization of standard model
f\/ields has been proposed, detail study of its implications is not yet
available. This review aims to survey the various ef\/forts and results.}


\Classification{83C45; 81T20; 81T25}

\renewcommand{\thefootnote}{\arabic{footnote}}
\setcounter{footnote}{0}

\section{Introduction}\label{Intro}

For most physical situations, we have the comfort of a background
space-time geometry -- be it the space and time of the non-relativistic
regime or the Minkowski space-time of special relativistic regimes or
even more exotic black hole space-times. Even to have a well def\/ined,
causally well behaved and deterministic classical theory of f\/ields, the
background space-times are limited to the class of {\em globally
hyperbolic} geometries. Quantum f\/ield theories can be def\/ined in all
such space-times \cite{WaldQFT}. However, the familiar Fock
representation or the particle interpretation can be def\/ined {\em
uniquely} only for a sub-class of space-times namely those that are {\em
stationary} or at least asymptotically (in future and past) stationary.
This is because in relativistic theories, the Fock representation
depends crucially on having a notion of positive (negative) energy
solutions, i.e.\ a notion of preferred time \cite{WaldCST}.

However, in regimes of early universe or evaporating black holes, we
loose this comfort of having a f\/ixed background space-time. In these
cases, not only do we have to face `quantum gravity' but also face the
issue of quantization of other matter quantum f\/ields {\em in the absence
of a background space-time}.

One strategy to construct quantum theories is to use the
Gelfand--Naimark--Segal theo\-ry~\cite{GNS}. Given a commutative $C^*$ algebra,
for every positive linear functional on it, there is a~cyclic, unitary
representation. The strategy is to look for a suitable sub-algebra of
the Poisson bracket algebra of classical observables, identify its
commutative sub-algebra and construct various representation.  Choose
among these, those on which the full sub-algebra is suitably
represented.  Under favourable conditions, a unique representation is
picked out. A well known example of this is the unique, weakly
continuous representation of the Weyl--Heisenberg $C^*$ algebra
(non-commutative)~\cite{WeylAlgebra,WeylAlgebra2}. The other example is the
holonomy-f\/lux representation of Loop Quantum Gravity (LQG)
\cite{UniqueLQG2,UniqueLQG}.

\looseness=1
A natural commutative $C^*$ algebra is constructed from suitable {\em
configuration space variables}. For a linear conf\/iguration space such as
$\mathbb{R}^{N}$, the coordinates themselves form suitable variables.
For a more general non-linear conf\/iguration space, coordinates have only
local meaning and are not suitable variables. Instead, one constructs a
set of {\em functions} on the conf\/iguration space which are invariant
under change of local coordinates. These will be more in number than the
dimension of the conf\/iguration space (manifold) and may possibly have
some relations among them. For example, for the conf\/iguration space
$S^2$, we can choose 3 functions on it -- namely $X^i(\theta, \phi)$, $i =
1, 2, 3$ satisfying $\sum_i (X^i)^2 = 1$ (say). The chosen class of
functions is required to constitute a {\em separating set of functions}
so that these functions serve to distinguish dif\/ferent points of the
conf\/iguration space. A set of functions $\{f_i\}$ on a space $Q$ forms a
separating set for~$Q$ if, for every pair of distinct points, there are
at-least two functions in the set which have distinct values at these
points.

For f\/ield theories, there is a further feature. The conf\/iguration space
is heuristically described in terms of (tensor etc) f\/ields on a
manifold. The `delta function' in the canonical brackets, means that the
canonical coordinates are to be understood as suitably smeared f\/ields.
For example, for scalar f\/ield $\phi$, $F_f(\phi) := \int_{\Sigma}d^3x
f(x) \phi(x)$, for all scalar densities of weight~1, $f$'s, def\/ines
functions on the conf\/iguration space. For a gauge f\/ield
$T_iA^i_{a}dx^{a}$, the holonomies $h_{e}(A) := {\rm Pexp}(\int_e A)$, is
another example of such functions.

It turns out that suitable representation of the Poisson bracket algebra
can also be constructed using a special class of functions on the
conf\/iguration space, the so called {\em cylindrical functions}, in
conjunction with projective techniques \cite{ALReview}. When applied to
formulation of gravity in terms of the real SU(2) connection, this
constructs the {\em kinematical Hilbert space} of LQG\footnote{The
cylindrical functions in this case are f\/inite linear combinations of
{\em spin network functions}. To def\/ine these, take an oriented,
piece-wise analytic, closed graph $\gamma$, embedded in the `spatial
manifold', $\Sigma_3$. Associate to each of its $E$ edges an $SU(2)$
group element in a representation $j$. These are the holonomies of a
(generalized) connection. Associate to each of its $V$ vertices, an
intertwiner $c$ and f\/inally take the trace (contract all indices). These
functions of connections are naturally labelled by $(\gamma, j_1,
\dots, j_E, c_1, \dots, c_V)$. Similar functions are def\/ined for
matter f\/ields as well.}. On this Hilbert space, geometrical operators
have discrete spectra and their eigenstates are associated with
appropriate class of graphs. This is paraphrased by saying that quantum
geometry lives on {\em graphs}. The quantum excitations of matter are
thus also expected to be associated with {\em vertices and edges} of
graphs.

Construction of such a representation constitutes only a f\/irst step in
the construction of a quantum theory of gravity. Due to the
dif\/feomorphism invariance, the system has {\em gauge invariances} and
the (Dirac) quantization procedure is split into {\em three} stages:
construction of the {\em kinematical Hilbert space} followed by
(distributional) solutions of the constraints followed by specif\/ication
of physical Hilbert space.

\looseness=1
In this review we will discuss the construction of kinematical Hilbert
space. Our discussion is conf\/ined to {\em four} space-time dimensions
wherein the matter f\/ields can be {\em scalars, spinors and $1$-form gauge
fields}. We will not discuss the gravitational sector (discussed in~\cite{ALReview} for instance) except to use the real SU(2) variables~--
connection and densitized triad~-- when needed in the matter sector.  Our
discussion is within a canonical framework leaving out discussion of
matter in a spin foam framework. Each type of matter is discussed in a
separate section. Section~\ref{Scalars} discusses loop quantization of
scalars and some of its implications.  Both cases of scalar f\/ields
ranging over a compact set (Higgs scalars in the adjoint representation
of a gauge group~\cite{ThiemannScalar}) and those ranging over the full
$\mathbb{R}$ \cite{BohrQuantized} are discussed.  Section~\ref{Fermions}
describes fermions, their Hamiltonian formulation and loop quantization
while Maxwell and Yang--Mills f\/ields are discussed in Section~\ref{YM}.
Propagation of matter waves on quantum geometry is brief\/ly discussed in
Section~\ref{MatterWaves}.  Finally, in Section \ref{Summary} we
make a few remarks.

\section{Scalars}\label{Scalars}

In loop quantum gravity approach, the scalar f\/ields, like other matter
f\/ields, remain comparatively under-studied. Often the scalar f\/ield has
been used to play a secondary role for describing dynamics of
gravitational degrees of freedom. In particular, it has been used as a
`clock' variable in loop quantum cosmology.  Nevertheless, there have
been some notable ef\/forts for studying quantum scalar f\/ields themselves.
In this section, we survey these dif\/ferent ef\/forts involving quantum
scalar f\/ields.

\subsection{Scalar f\/ield as a `clock'}\label{ScalarClock}

We begin the review for scalar f\/ields by considering the situations
where scalar f\/ields have been used as `clock' variables. In these
studies, mostly for cosmological scenario, scalar f\/ields have been
quantized using standard quantization. Of course, one should use loop
quantization also for scalar f\/ields, as done for gravitational degrees
of freedom.

In a general background the equation of motion of a \emph{massless,
free} scalar f\/ield $\phi$ is
\begin{gather}
\label{ScalarLEoMGeneral}
\partial_{\mu} \big( \sqrt{-g} g^{\mu\nu} \partial_{\nu}\phi  \big) =
0 ,
\end{gather}
where $g$ is the determinant of the metric $g_{\mu\nu}$. For
cosmological backgrounds, i.e.\ in a \emph{spatially homogeneous}
space-time, the equation of motion~(\ref{ScalarLEoMGeneral}) reduces to
\begin{gather}
\label{ScalarLEoMCosmo}
\partial_{t} \big( \sqrt{q(t)}  N^{-1} \partial_{t} \phi \big) = 0.
\end{gather}
Now if one makes the choice of lapse function $N$ to be $\sqrt{q(t)}$
where $q$ is the determinant of the spatial metric then the equation~(\ref{ScalarLEoMCosmo}) implies scalar f\/ield $\phi \propto t$.  This
suggests that one may choose a massless scalar f\/ield as an `internal
clock' to describe the dynamics of the remaining degrees of freedom.
Use of `internal time' to describe the dynamics, instead of the
coordinate time~$t$, has the advantage of avoiding the so-called
`problem of time' in quantum gravity.

By choosing such a lapse function, the total Hamiltonian density for a
cosmological system with a massless free scalar f\/ield can be written as
($N = \sqrt{q}$ chosen)
\begin{gather*}
H = H_{g} +  \frac{1}{2} \pi_{\phi}^2 ,
\end{gather*}
where $H_{g}$ represents gravitational Hamiltonian. Now imposition of
Hamiltonian constraint on the Hilbert space, i.e.\ $\hat{H} \Psi =
0$ implies
\begin{gather*}
\partial_{\phi}^2 \Psi  =  2 \hat{H}_{g} \Psi,
\end{gather*}
where we have used $\hat{\pi}_{\phi} =  -i \partial_{\phi}$.  Thus, the
positive and negative frequency solutions can be viewed as satisfying a
set of f\/irst order evolution equations of the form
\begin{gather*}
\pm i \partial_{\phi} \Psi = \sqrt{-2 \hat{H}_{g}}  \Psi  =:
\hat{\mathbb{H}}  \Psi .
\end{gather*}
By identifying the f\/ield $\phi$ to play the role of time and
$\hat{\mathbb{H}}$ that of a proper Hamiltonian,  one may note that the
positive or negative frequency solutions satisfy appropriate
Schr\"odinger equation with respect to the internal time $\phi$. This
approach of considering scalar f\/ield as an internal time has been used
to describe the quantum dynamics of isotropic cosmological models such
as Friedmann--Robertson--Walker space-time \cite{APS:QuanBBAnaNum} as well
as anisotropic cosmology such as Bianchi~I space-time \cite{AW:LQCBI}.
It has been shown for isotropic models that universe undergoes a
\emph{bounce}  \cite{APS:Improved} when quantum dynamics is described
with respect to an internal time.

\subsection{Compact scalars}\label{Compact}
The Higgs scalars plays a very important role in standard model of
particle physics. It provides a mechanism for generation of masses for
dif\/ferent elementary particles. Naturally, in any approach to quantum
gravity, one would  expect to have an appropriate representation of the
Higgs scalars in the given quantum theory.

In the standard model of particle physics, while constructing the
quantum framework one specif\/ically uses the fact that the background
space-time is a Minkowski space-time. A natural kinematical measure
available in constructing a suitable Hilbert space in a such space-time
is Gaussian measure. Use of this measure leads to the standard Fock
Hilbert space. However, one cannot use  Gaussian measure in a background
independent construction of the Hilbert space such as those used in loop
quantum gravity.  This is because Gaussian measure is not dif\/feomorphism
invariant.

In order to avoid this problem, Thiemann proposes \cite{ThiemannScalar,ThiemannQSD,ThiemannQSD2,ThiemannQSD4} to
use a set of bounded variables for quantizing Higgs scalars. In
particular, bounded variables of the form $U(x) = e^{\phi_I(x) \tau^I}$
are used as the conf\/iguration space variables rather than the scalars
$\phi_I$ themselves. Here, scalars $\phi_I$ have been made dimensionless
with the help of a constant scale of $\sqrt{8\pi G}$ where $G$ is the
Newton's constant of gravitation.  These variables are the analogues of
the holonomies of connection variables. Unlike the gauge holonomies
which are labelled by curves, these are labelled by {\em points} (no
smearing) and are referred to as `point-holonomies'. These are valued in
$\mathcal G$ since the $\phi_I(x)\tau^I$ are valued in its Lie algebra.  Note
that $U(x)$ is a matrix representing a {\em group element} in a certain
representation determined by the generators $\tau^I$. Matrix elements of
$U(x)$ therefore determine functions of the Higgs scalars. If one
considers a single real scalar f\/ield, valued in a~compact interval $[0,
2\pi]$ (say), then the basic variables can be chosen to be $U(x) =
e^{i\phi(x)}$ which are valued in $U(1)$. For constructing the
kinematical Hilbert space, one considers products of f\/initely many such
matrix elements labelled by a f\/inite set of points of the spatial
manifold together with unitary representations of $\mathcal G$ associated with
them. Finite linear combinations of these functions generate the
(quantum) conf\/iguration space $\overline{\mathcal{U}}$ and the
kinematical Hilbert space is then obtained as the space of square
integrable functions $L_2(\overline{\mathcal{U}}, d\mu_H)$ where
$d\mu_H$ is naturally available Haar measure on the group~$\mathcal G$. On this,
the point holonomies, but not the scalar f\/ield, are well def\/ined
(multiplicative) operators. The momentum is a density weight~1 object
and its integral over three-dimensional regions are analogues of the
f\/lux variables of the gravity sector. Their action on the polynomials
needs careful def\/inition for which we refer to~\cite{ThiemannScalar}. All other operators of interest are to be
constructed using these basic operators.

We may recall that the classical dynamics of a real scalar f\/ield is
governed by the Hamiltonian
\begin{gather}
\label{ScalarHamiltonian}
H_{\phi} =  \int d^3x N  \left[ \frac{\pi_{\phi}^2}{2\sqrt{q}}   +
\frac{1}{2}  \sqrt{q}  q^{ab} \nabla_a \phi \nabla_b \phi +
\sqrt{q}V(\phi)  \right].
\end{gather}
Formally, one can construct the quantum Hamiltonian operator
corresponding to the classical Hamiltonian~(\ref{ScalarHamiltonian}).
Regularization of this quantum Hamiltonian operator is performed by
using the triangulation of manifold which is analogous to the
construction of gravity sector operators. While formal construction of
the scalar Hamiltonian has been performed by Thiemann, understanding
it's physical consequences remains an open issue.

\subsection{Non-compact scalars}\label{Noncompact}

In Thiemann's construction \cite{ThiemannScalar}, the conf\/iguration
space variables are essentially \emph{periodic} functions of the scalar
f\/ield. This is suf\/f\/icient to describe the scalar f\/ields which range over
a compact interval. However, such treatment of scalars is inadequate to
describe the Klein--Gordon type real-valued scalar f\/ield which takes
values in $\mathbb{R}^1$ rather than in $U(1)$. To circumvent this
problem, Ashtekar, Lewandowski and Sahlmann propose  the use of
\emph{almost periodic} functions as the conf\/iguration space variables~\cite{BohrQuantized}. The formal construction of the kinematical Hilbert
space for a single non-compact scalar is carried out as follows: First
one considers f\/inite sets of points in~$\mathbb{R}^3$ referred to as the
vertex sets. For a given vertex set $V = \{x_1, \dots, x_n\}$, the
corresponding vector space is generated by linear combinations of the
functions of the form
\begin{gather*}
\psi_{V, \vec{\lambda}}(\phi) = e^{i \sum_j \lambda_j \phi(x_j)} ,
\end{gather*}
where $\lambda_j$'s are real numbers.  After considering all possible
sets of vertices, the union of corresponding vector spaces and together
with its completion, one construct the quantum conf\/iguration space
required for loop representation. If one denotes this space as
$\overline{\mathcal{A}}$ then the corresponding Hilbert space is given
by $L_2(\overline{\mathcal{A}}, d\mu)$. The measure $d\mu$ is def\/ined as
follows
\begin{gather}
\label{KGmeasure}
\int_{\bar{\mathcal{A}}} d\mu \psi_{V, \vec{\lambda}} =
\begin{cases}
1,  &  \text{if $\lambda_j = 0 \  \forall \, j$,}  \\
0,   &    \text{otherwise.}
\end{cases}
\end{gather}
The elementary conf\/iguration operators are analogous of the holonomy
operator for connection variables and they act by multiplications as
\begin{gather*}
\hat{h}(x,\lambda) \Psi = e^{i\lambda \phi(x)} \Psi .
\end{gather*}
Use of the measure (\ref{KGmeasure}), makes the elementary holonomy
operator $h(x,\lambda) = e^{i\lambda\phi(x)}$ \emph{not} to be weakly
continuous in the parameter $\lambda$. This implies that one cannot
construct an operator corresponding to the scalar f\/ield $\phi(x)$
itself. The conjugate variable to the holonomy which is promoted as an
operator in the quantum theory, is taken to be the smeared f\/ield
momentum
\begin{gather*}
P(f) = \int dx^3 \pi_{\phi}(x) f(x) ,
\end{gather*}
where $\pi_{\phi}$ is conjugate momentum f\/ield and  $f(x)$ is a test
function. Their Poisson bracket is given by
\begin{gather}
\label{KGPBSmeared}
\{h(x,\lambda), P(f) \} = i\lambda f(x) h(x,\lambda).
\end{gather}
Subsequently, one looks for a representation of the Poisson bracket~(\ref{KGPBSmeared}) in the quantum theo\-ry as a commutator of the
elementary operators. The  non-trivial commutator bracket of the
elementary operators is
\begin{gather*}
[ \hat{h}(x,\lambda), \hat{P}(f) ] = i\lambda f(x) \hat{h}(x,\lambda),
\end{gather*}
as holonomy operators and smeared momentum operators commute with
themselves. After having the kinematical framework ready, one may
construct the scalar Hamiltonian operator to understand its quantum
dynamics.

\subsection{Parametrized scalar f\/ields}\label{PFT}

Within loop approach, scalar f\/ields have been studied in the context of
parametrized f\/ield theory (PFT). In the PFT approach, scalar f\/ield
theory on a f\/lat space-time is recast in the form as if the f\/ields
reside in a generally covariant background. In the references
\cite{LM:PFT1,LM:PFT2}, Laddha and Varadarajan have studied parametrized
f\/ield theory for a free scalar f\/ield $\phi$ in $2$-dimensional f\/lat
space-time. The action for such a scalar f\/ield with respect to an
inertial frame is
\begin{gather} \label{LVPFTAction0}
S[{\phi}] = \int
d^2X \eta^{AB} \partial_A \phi \partial_B \phi ,
\end{gather} where
metric $\eta_{AB} = {\rm diag}(-1,1)$ and $A=0,1$. Now one may choose to use
an arbitrary coordinate system involving $x^a$ with $a=0,1$ such that
$X^A$ can be viewed as `parametrized' by~$x^a$ as  $X^A = X^A (x^a)$.
Then the action (\ref{LVPFTAction0}) can be rewritten as
\begin{gather*} 
S[{\phi}] = \int d^2x  \sqrt{g}
g^{ab} \partial_a \phi \partial_a \phi ,
\end{gather*}
 where $g_{ab}
= \eta_{ab} \partial_a X^A \partial_b X^B$ and $g$ is the absolute value
of the determinant of the metric $g_{ab}$.

In the parametrized f\/ield theory, the action $S[\phi]$ is treated as not
only a functional of $\phi$, but also a functional of $X^A$'s. In other
words the action for parametrized f\/ield theory is taken to be
\begin{gather}
\label{LVPFTAction2}
S[{\phi}, X^A] = \int d^2x \sqrt{g(X)} g^{ab}(X) \partial_a \phi
\partial_a \phi  .
\end{gather}
The variation of the action (\ref{LVPFTAction2}) with respect to the
f\/ield $\phi$ leads to $\partial_a  (\sqrt{g} g^{ab} \phi \partial_a
\phi  ) = 0$. This equation is basically the same as the f\/lat space
equation of motion for $\phi$, i.e.\ $ \eta^{AB} \partial_A
\partial_B \phi=0$ but written in~$x^a$ coordinates.  The variation with
respect to $X^A$ leads to the equations which are automatically
satisf\/ied as long as $ \eta^{AB} \partial_A \partial_B \phi=0$. In other
words, choice of the new coordinates remains undetermined as expected.

For canonical formulation of this theory, it is convenient to choose the
light-cone coordinates def\/ined as $X^{\pm} \equiv X^0 \pm X^1$.  One may
denote the corresponding conjugate momenta by $\Pi_{\pm}$. In terms of
these coordinates the action~(\ref{LVPFTAction2}) can be expressed as
\begin{gather*}
S = \int dt \int dx \big[ \pi_{\phi} {\dot \phi}  + \Pi_+ {\dot X^+} +
\Pi_- {\dot X^-}  - N^+ H_+ - N^- H_- \big ]  ,
\end{gather*}
where $N^{\pm}$ are the Lagrange multipliers for the respective
constraints~$H_{\pm}$. As constructed, the non-trivial Poisson brackets
are
\begin{gather} \label{LMPFTPB}
\{\phi(x), \pi_{\phi}(x')\}  = \delta(x,x')   ,  \{X^{\pm}(x),
\Pi_{\pm}(x')\}  = \delta(x,x')  .
\end{gather}
In the standard Fock quantization, one seeks an appropriate
representation of the Poisson brac\-kets (\ref{LMPFTPB}) as commutator
brackets of the operators in the quantum theory.

To proceed to polymer quantization, one notes that in 1-space dimension
any tensor can be viewed as a density of weight equal to its covariant
rank minus its contravariant rank. Conversely, a density weight 1 object
can be viewed as a 1-form. Hence $\Pi_{\pm}, Y_{\pm} := \pi_f \pm
\partial_x\phi$ are all 1-forms which are integrated along intervals.
Exponentials of the integrals are the holonomies ({\em not point
holonomies}). The polymer quantization of the system consists of
choosing holonomies of~$\Pi_{\pm}$ as multiplicative operators for the
$X^{\pm}$, $\Pi_{\pm}$ sector  and choosing  holonomies of both~$Y_{\pm}$
in the matter sector as the basic variables (non-commuting). Details
should be seen in the references cited above.

Using this quantization it is shown that one can construct a suitable
state in the polymer Hilbert space which reproduces the Fock space two
points function in the large wavelength limit. However, understanding
the behaviour of two-point function in the strong polymer regime remains
an open issue.

\subsection{Polymer quantized scalars in Fourier space}\label{PolymerScalar}

A dif\/ferent study of scalar f\/ields using loop quantum gravity techniques
has done by Hossain, Husain and Seahra \cite{HHS:Propagator}. In their
approach,  polymer quantization is performed for the Fourier modes of
the free scalar f\/ield in f\/lat space-time. This has the advantage of
dealing with a known Hamiltonian which is a sum of decoupled harmonic
oscillators. One then computes $2$-point correlation function directly
in the energy eigenfunction basis.

The phase space variables for the free scalar f\/ield are the canonical
pair ($\phi(\x,t), \pi(\x,t)$) satisfying the Poisson bracket
\begin{gather*}
\{\phi(t,\x), \pi(t,\y)\} = \delta^{(3)}(\x-\y).
\end{gather*}
The Hamiltonian is
\begin{gather}
\label{SFHamGen}
H_{\phi}  =  \int d^3\x \left[ \frac{\pi^2}{2} + \frac{1}{2} \eta^{ab}
\partial_a\phi \partial_b\phi \right],
\end{gather}
where the space-time metric is $ds^2 = -dt^2 + \eta_{ab}dx^adx^b$.  It
is convenient though not essential to put the system in a box by
restricting to a f\/inite region of f\/lat $3$-space so that the volume $V =
\int d^3x \sqrt{\eta}$ is f\/inite.  One can then perform Fourier
expansion of the f\/ield into the $3$-momentum space as
\begin{gather*}
\phi(t,\x)  =  \frac{1}{\sqrt{V}} \sum_{\k}{\phi}_{\k}(t) e^{i
{\k}\cdot{\x}}, \qquad
{\phi}_\k(t)  =  \frac1{\sqrt{V}}\int d^3x\ e^{-i\k\cdot x} \phi(\x,t),
\end{gather*}
along with a similar expansion for $\pi(\x,t)$.  After a suitable
redef\/inition of the Fourier modes, one can express  scalar f\/ield
Hamiltonian (\ref{SFHamGen}) in terms of  \emph{real-valued} $\phi_{\k}$
and $\pi_{\k}$ as
\begin{gather*}
H_{\phi} =  \sum_{\k} H_{\k} = \sum_{\k} \left[ \frac{\pi_{\k}^2}{2} +
\frac{1}{2} k^2 \phi_{\k}^2 \right],
\end{gather*}
with the Poisson bracket being $\{\phi_\k, \pi_{\k'}\} =
\delta_{\k\k'}$.  Clearly, the Hamiltonian is a sum of \emph{decoupled}
harmonic oscillators.  The  polymer quantization of each mode then
follows that of a quantum oscillator.  In particular,  the variables
used in polymer quantization are  $\phi_\k$ and $U_{\lambda\k }=
e^{i\lambda \pi_k}$ which satisfy the  Poisson bracket $\{ \phi_\k,
U_{\lambda\k}\} = i \lambda U_{\lambda\k }$. The parameter $\lambda$ has
dimensions of $(\text{length})^{1/2}$.

As in the case of polymer quantization, $\pi_k$ itself cannot be a
well-def\/ined operator in the quantum theory as the action of
$U_{\lambda\k}$ is not \emph{weakly} continuous with respect to
$\lambda$. To represent momentum operator, instead one uses
$\pi_k^{\star} = (U_{\lambda_{\star}\k } - U_{\lambda_{\star}\k
}^{\dagger})/2i\lambda_{\star}$.  It may be emphasized that the polymer
quantization method introduces a new length scale (which is
$\lambda_{\star}$ here) in addition to Planck's constant into the
quantum theory. One can def\/ine a dimensionless parameter as
\begin{gather*}
g=\lambda_\star^2 |\k|  \equiv \frac{|\k|}{M_\star} \propto
\frac{\text{polymer length scale}}{\text{spatial wavelength}},
\end{gather*}
where $M_{\star}^{-1}$ is the fundamental length scale associated with
the polymer quantization of $\phi$. Clearly, $g=0$ should recover the
results that one get from Fock quantization. The standard  $2$-point
correlation function is def\/ined as
\begin{gather*}
\langle 0| \hat{\phi}(\x,t) \hat{\phi}(\x',t')|0\rangle \equiv
\frac{1}{V} \sum_{\k} e^{i {\k}\cdot(\x-\x')}D_{\k}(t-t') ,
\end{gather*}
where $|0\rangle=\Pi_{\k}\otimes |0_\k\rangle$ is the vacuum state and
the matrix element is
\begin{gather}\label{pprop-def}
D_{\k}(t-t') = \langle 0_{\k}| e^{i\hat{H}_{\k}t} \hat{\phi}_{\k}
e^{-i\hat{H}_{\k}t} e^{i\hat{H}_{\k}t'} \hat{\phi}_{\k}
e^{-i\hat{H}_{\k}t'} |0_{\k}\rangle,
\end{gather}
where $\hat{H}_\k$ is the Hamiltonian operator.  The matrix element
(\ref{pprop-def}) can be computed using  the polymer oscillator spectrum
$\hat{H}_{\k}|n_{\k}\rangle = E_n^{(\k)}|n_{\k}\rangle$,  and the
expansion of the state $\hat{\phi}_{\k} |0_{\k}\rangle$ in the energy
eigenstates as $\hat{\phi}_{\k} |0_{\k}\rangle = \sum_{n} c_n
|n_{\k}\rangle$. By def\/ining $4$-momentum $p \equiv(\omega,\k)$, one can
write the momentum space  propagator as
\begin{gather}\label{KGPropagatorMomentum}
D_{p} = \sum_{n} \frac{2i\Delta E_n |c_n|^2 }{p^2 + \Delta E_n^2 -
|\k|^2 - i\epsilon},
\end{gather}
where $p^2=-\omega^2 +|\k|^2$  and choice of the sign of $i\epsilon$
corresponds to the Feynman propagator.  One can recover the expected
Fock space result  $D_{p} = i/(p^2 - i\epsilon)$ by using properties of
the Schr\"odinger oscillator, i.e.\ $c_n =
\delta_{1,n}/\sqrt{2|\k|}$ and $\Delta E_n = n |\k|$.
The propagator (\ref{KGPropagatorMomentum}) with polymer corrections in
the \emph{infrared limit} ($g\ll 1$) can be  written as
\begin{gather*}
D_p = \frac{i(1-2g) }{p^2   - g|\k|^2 - i\epsilon }  + {\cal O}\big(g^2\big).
\end{gather*}
The pole of the propagator implies an ef\/fective dispersion relation
\begin{gather*}
\omega^2 =|\k|^2(1 -|\k|/M_\star),
\end{gather*} which violates Lorentz symmetry. On the other hand in the
\emph{ultraviolet limit} ($g \gg 1$), the propagator with leading order
corrections is
\begin{gather*}
D_p =  \frac{i/8g^2}{p^2 + 4 g^2|\k|^2-i\epsilon} +{\cal
O}\left(\frac{1}{g^6}\right).
\end{gather*}
The corresponding dispersion relation
\begin{gather*}
\omega^2 = 4|\k|^4/M_\star^2
\end{gather*}
also violates Lorentz invariance. One may also note that the propagation
amplitude at high momentum is suppressed by a factor $1/g^2$.

Polymer quantization introduces a length scale $\lambda^2 = M_*^{-1}$.
This is unavoidable. Such a~scale could be provided by an underlying
quantum geometry. However at the present level of understanding, it is
to be treated as an arbitrary parameter. Current observational limits on
Lorentz violation may be then used to constrain its value.

\subsection{Other studies of scalar f\/ields}\label{ScalarApp}

Ultraviolet behaviour of polymer quantized scalar f\/ields has been
studied by Husain and Kreienbuehl by constructing Fock-like states in
f\/lat space-time \cite{Husain:UVBehavior}. Instead of Fourier space, as
discussed in section (\ref{PolymerScalar}), here one applies polymer
quantization in real space.  It is shown that the vacuum expectation
values of the commutator and anti-commutator of the creation and
annihilation operators become energy dependent, and tends to show some
sort of fermionic behavior at high energy. Furthermore, the modif\/ied
dispersion relation that arises leads to violation of Lorentz
invariance.

In the context of spherically symmetry, Gambini, Pullin, and Rastgoo
have studied polymer quantization of scalar f\/ield \cite{Gambini:2011mw}
coupled to gravity. Their study is performed in the midi-superspace
context where even after imposition of the given symmetry, f\/ield
theoretical nature of scalar f\/ields, i.e.\ with inf\/initely many
degrees of freedom, is retained. After choosing a~quantum geometry state
corresponding to locally f\/lat geometry, quantization of the spherically
symmetric scalar f\/ield is carried out using a discretization followed by
a `polymerization' done in two dif\/ferent ways (the f\/ield and the
momentum). Their notable conclusion is that the propagator obtained is
not Lorentz invariant.

Polymer quantized free massless scalar f\/ield in a homogeneous and
isotropic cosmological space-time has been studied by Hossain, Husain
and Seahra~\cite{HHS:CosmoPM}. Use of semi-classical Friedman equation
yields a non-singular and non-bouncing universe, without quantum
gravity. The system exhibits an early de Sitter-like inf\/lationary phase
with suf\/f\/icient expansion to resolve the horizon and entropy problems,
and a built in mechanism for a graceful exit from inf\/lation.

The propagation of polymer quantized scalar f\/ield in f\/lat space-times
has been studied in~\cite{HHS:WaveProp}.  Polymer quantization is
performed in real space there unlike in Fourier space as discussed in
section (\ref{PolymerScalar}).  Using semi-classical states, it has been
shown that ef\/fective wave equation is both nonlinear and Lorentz
invariance violating. Furthermore, it is demonstrated that polymer
ef\/fects tend to accumulate with time for plane-symmetric waveforms.
Possibility of measuring deviations from the Klein--Gordon equation in
particle accelerators or astrophysical observations is also discussed.

\section{Fermions}\label{Fermions}
Fermions coupled to gravity has been discussed in the literature
\cite{FermionHistory7,FermionHistory2,FermionHistory3,FermionHistory4,FermionHistory5,FermionHistory6,FermionHistory}, at both classical and quantum level. In the
formulation of general relativity in terms of real SU(2) connection,
Thiemann discussed loop quantization of standard model f\/ields
\cite{ThiemannScalar,ThiemannQSD,ThiemannQSD2,ThiemannQSD4}.  The fermions were treated in the second order form,
i.e.\ fermions couple to gravity through the spin connection (torsion
free Lorentz connection). Perez and Rovelli returned to fermions in
presence of the Holst term and found that the Barbero--Immirzi parameter,
$\gamma$, inverse of the coef\/f\/icient of the Holst term, becomes
classically observable~\cite{PerezRovelli}.  Mercuri~\cite{Mercuri,Mercuri2}
discovered that with a further addition of suitable non-minimal
fermionic couplings, $\gamma$ can be made classically unobservable. He
also noted that the added terms (Holst plus non-minimal) can be
expressed as the Nieh--Yan topological term after using the connection
equation of motion.  This strategy of adding non-minimal couplings to
keep $\gamma$ classically unobservable was followed for $N = 1, 2$ and~4
supergravities also \cite{Kaul}. Ca\-no\-ni\-cal analysis and loop
quantization of fermions with non-minimal couplings was discussed by
Bojowald and Das~\cite{BojowaldDas,BojowaldDas2}.

It was subsequently realised that $\gamma$ will automatically be
classically unobservable provided it is the (inverse) of the coef\/f\/icient
of the Nieh--Yan term (a total divergence) in the Lagrangian density.
Thus, instead of the Holst terms alone, if the Nieh--Yan term (Holst +
(torsion)$^2$ piece) is used in conjunction with the Hilbert--Palatini,
then for {\em arbitrary} matter and their couplings, the $\gamma$ will
drop out of the classical equations of motion\footnote{A necessary
condition for a topological origin of $\gamma$ is thus satisf\/ied.}.  But
now, since the action is modif\/ied, it was not obvious that the real
SU(2) formulation will result from the new action. It turned out that it
is possible to systematically derive the real SU(2) Hamiltonian
formulation from such an action \cite{DateKaulSengupta,
KaulSenguptaSugra}.  Since it is in presence of fermions that
non-trivial torsion results from the equation of motion of the Lorentz
connection, fermions were also included in the canonical analysis and
real SU(2) formulation was seen to emerge.  The canonical analysis
leading to real SU(2) formulation has since been extended to include the
other two topological terms namely the Pontryagin and the Euler classes
\cite{KaulSenguptaTop,PerezRezende}.

It is straight forward to derive the real SU(2) formulation from the
Hilbert--Palatini action with $\gamma^{-1}$ times the Nieh--Yan term
\cite{DateNew,DateKaulSengupta}. Its main points may be summarised as
follows. In the {\em time gauge parametrization}\footnote{$I, J = 0, 1,
2, 3$ are the Lorentz indices, $a = 1, 2, 3$ denote the spacial indices.
The co-tetrad components are $e^0_t := N$, $e^i_t := N^aV^i_a$, $e^i_a := V^i_a$, $q_{ab} := V^i_a V^j_b \delta_{ij}$, $q :=
\det(q_{ab})  \neq 0 $. The Lorentz connection components are
denoted as $\omega^{0i}_a =: K^i_a$, $\omega_a^{ij} =: {\cal
E}^{ij}_{~\ k} \Gamma^k_a$~\cite{DateNew}.}, to begin with there are
13 components of the co-tetrad  ($N, N^a, V_a^i$) and 24 components of
the Lorentz connection ($\omega_t^{IJ}, \omega_a^{IJ}$).  Of these, 10
variables~-- $N$, $N^a$, $\omega_t^{IJ}$~-- occur as Lagrange multipliers and
18 variables~-- $E^a_i \sim V^a_i$ (densitized triad) and 9 combinations
$A^i_a$ of $\omega_a^{IJ}$~-- explicitly appear in the form $p\dot{q}$.
The coef\/f\/icients of the Lagrange multipliers form the primary
constraints ${\cal H}_a$, ${\cal H}$, ${\cal G}^{0i}$ and ${\cal G}^{ij}$.
The remaining 9 combinations of the Lorentz connections, have no
velocities and lead to additional 9 primary constraints, say $\pi_i
\approx 0 \approx \pi^{ij}$ (symmetric in~$i$,~$j$). Of these, the 3 primary
constraints, $\pi_i \approx 0$ form a~second class system with the
primary constraints ${\cal G}^{0i}$ and are eliminated easily using
Dirac brackets. Preservation of the remaining~6 primary constraints
$\pi^{ij} \approx 0$ lead to further 6 secondary constraints~$S_{ij}
\approx 0$, with which they form a second class system. These are again
eliminated using Dirac brackets.  The left over system has the 18 phase
space variables, ($A^i_a, E^a_i$), and~7~f\/irst class constraints and is
the real SU(2) connection formulation. This counting and the steps in
the constraint analysis remain the same when matter is included.

This derivation also introduces factors of $\sgn :=
 {\rm sign}(\det(e^I_{\mu}$)) $(= N\,{\rm sign}(\det(V^i_a))$ in the time gauge
parametrization) in appropriate places. For instance, we are naturally
lead to the def\/initions: $E^a_i :=  \sgn \sqrt{q}V^a_i $ and $ A^i_a :=
\gamma^{-1} \sgn\, K^i_a -  \Gamma^i_a$ Under an {\em improper} orthogonal
transformation $\Lambda^i_{~j} \in {\rm O}(3)$ acting on the index $i$, the
triad changes its handedness and the $\sgn$ factor changes sign leaving
the handedness of $E^a_i$ unchanged.  This is as it should be since the
index $i$ on $E$ represents adjoint representation of SO(3) while on~$V$
it represents the def\/ining representation. For SO(3), both are
equivalent but {\em not for} O(3).  Under an {\em inversion},
$\Lambda^i_{~j} = - \delta^i_{~j}$, quantities in the def\/ining
representation change sign while those in the adjoint don't. The same
reasoning applies to the def\/inition of the connection.  Now the
connection is also even under inversion. The $\sgn$ factor also change
the behaviour of $E^a_i$ and $A^i_a$ under the action of {\em
orientation reversing dif\/feomorphisms}. These factors are relevant for
discussion of `parity' properties of the canonical formulation
(discussed later). This is independent of matter couplings.

When a Dirac fermion is included, the solution of the secondary, second
class constraint $S_{ij} \approx 0$ changes since fermions couple to
Lorentz connection apart from coupling to the triad. This solution leads
to non-trivial Dirac brackets between the SU(2) connection and the
fermions. One can however make natural shifts in the def\/inition of the
connection to recover the canonical brackets.  This also simplif\/ies the
constraints.  Four Fermi interaction terms however survive in the
Hamiltonian and are signatures of f\/irst order formulation. In the second
order formulation where fermions couple to the torsion free connection,
there are no terms quartic in the fermions.

Fermions are also tied with possible parity violations \cite{BojowaldDas,BojowaldDas2, Freidel}.  There are two distinct notions of `parity': one related
to orientation of the space-time manifold ({\em parity}) and one related
to the improper Lorentz transformation ({\em Lorentz parity}). Depending
upon the def\/initions of the basic canonical variables (with or without
the $\sgn$ factors in this work (Section~\ref{Parities})), the canonical
framework and the action are (non-)invariant under {\em one} of the
notions of parity. These possibilities are distinguished and discussed
in \cite{DateNew}.  Bojowald and Das discuss the non-invariance under
{\em Lorentz parity} \cite{BojowaldDas,BojowaldDas2} in the context of non-minimally
coupled fermion.

We now turn to the classical, canonical form of a Dirac fermion
minimally coupled to gravity in the f\/irst order formulation and discuss
the loop quantization of fermions in the following sub-section.

\subsection{The Hamiltonian formulation}\label{CanonicalFormulation}

\looseness=1
The starting point is a choice of (tensor/spinor) f\/ields and a
corresponding generally covariant, local action on 4-dimensional
space-time $M \simeq \mathbb{R} \times \Sigma_3$. The next step is to
carry out a 3+1 decomposition to identify the Lagrangian which is a
function of (tensor) f\/ields on $\Sigma_3$ together with their velo\-ci\-ties
with respect to the chosen time coordinate. The f\/ields whose velocities
appear in the Lagrangian are {\em potentially} the conf\/iguration space
variables while those without velocities appearing in the Lagrangian are
{\em Lagrange multipliers} whose coef\/f\/icients will be {\em primary
constraints}. This Lagrangian leads to the {\em kinematical phase
space}. Now a constraint analysis a la Dirac is performed. If there are
second class constraints, one may hope to simplify the analysis by
solving the second class constraints. However, now one must use the
Dirac brackets. These may not have the canonical form for the remaining
variables (i.e.\ may not be Darboux coordinates) and a new choice of
variables may be necessary.  This is particularly relevant for
Lagrangians which are {\em linear } in the velocities eg the
Hilbert--Palatini--Nieh--Yan and the Dirac Lagrangians which typically do
have second class constraints. The classical Hamiltonian formulation is
completed when the action is expressed in the Hamiltonian form together
with f\/irst class constraints.  We also have f\/ields coordinatizing the
kinematical phase space (after the second class constraints are
eliminated) with the conf\/iguration space coordinates identif\/ied.

We begin with the Lagrangian 4-forms built from the basic f\/ields the
co-tetrad $e^I_{\mu}dx^{\mu}$ and the Lorentz connection
$\omega^{IJ}_{\mu}dx^{\mu}$ $(\kappa := 8\pi G)$
\begin{gather*}
{\cal L}_{\mathrm{HP}}(e, \omega)  =  \frac{1}{2\kappa}\left[
\sgn \frac{1}{2}{\cal E}_{IJKL}R^{IJ}(\omega)\wedge e^K
\wedge e^L\right], 
\nonumber\\
{\cal L}_{\mathrm{NY}}(e, \omega)  =  \left[T^I(e, \omega)\wedge
T_I(e, \omega) - R_{IJ}(\omega)\wedge e^I\wedge e^J\right], 
\nonumber\\
{\cal L}_{\mathrm{grav}}  :=  {\cal L}_{\mathrm{HP}} +
\frac{\eta}{2\kappa}{\cal L}_{\mathrm{NY}} , 
\nonumber\\
{\cal L}_{\mathrm{Dirac}}  =  - \frac{i}{2} |e| \left[ \bar{\lambda}
e^{\mu}_I \gamma^I D_{\mu}(\omega, A, \ldots) \lambda -
{\overline{D_{\mu}(\omega, A, \ldots) \lambda}} e^{\mu}_I \gamma^I
\lambda \right], \\
D_{\mu}(\omega)\lambda  :=  \partial_{\mu} \lambda +
\frac{1}{2}\omega_{\mu}^{\ IJ} \sigma_{IJ}\lambda + i e'A_{\mu}\lambda +
\dots+ \lambda, \nonumber \\
\overline{D_{\mu}(\omega)\lambda}  :=  \left\{\partial_{\mu}
\lambda^{\dagger} + \frac{1}{2}\omega_{\mu}^{\ IJ}
\lambda^{\dagger}\sigma^{\dagger}_{IJ} - i e' A_{\mu} \lambda^{\dagger}
+ \dots + \lambda^{\dagger} \right\}\gamma^0.\nonumber
\end{gather*}
The $\ldots$ refer to possible couplings of the Dirac fermion to other
gauge f\/ields, e.g.~the Maxwell f\/ield. These are suppressed in the
following\footnote{Our conventions are: Greek letters denote space-time
indices, lower case Latin letters denote space indices, upper case Latin
ones denote Lorentz indices. Our signature is $(- + + +)$. The (metric
independent) Levi-Civita symbols are ${\cal E}^{txyz} = +1 = {\cal
E}_{0123}$. The determinant is def\/ined by $ e\ {\cal E}^{IJKL}   =   -
{\cal E}^{\mu\nu\alpha\beta}e^I_{\mu} e^J_{\nu} e^K_{\alpha}
e^L_{\beta}.$
For the Dirac matrices:
$2\eta^{IJ}\mathbb{1}  =  \gamma^I\gamma^J + \gamma^J\gamma^I$, $\sigma^{IJ}  := \frac{1}{4}\left[ \gamma^I,
\gamma^J\right]$, $\gamma_5  := i\gamma^0\gamma^1\gamma^2\gamma^3$, $\bar{\lambda}   :=  \lambda^{\dagger}\gamma^0.$\label{Notation}}.

Here $R := d\omega + \omega\wedge\omega$ and $T := de + \omega\wedge e$
are the usual curvature and torsion 2-forms.  The factor of sgn($e$) is
present because only then the Hilbert--Palatini Lagrangian matches with
the $\sqrt{|g|} R(g)$. This arises from noting that determinant of the
co-tetrad is given by, $e = \sgn  \sqrt{|g|}$.

A 3+1 decomposition is carried out as usual by choosing a foliation
def\/ined by {\em a time function}, ${\cal T}:M \to \mathbb{R}$ and a
vector f\/ield $t^{\mu}\partial_{\mu}$, transversal to its leaves. The
vector f\/ield is normalised by $t\cdot\partial {\cal T} = 1$ so that the
parameters of its integral curves, serve as the time coordinate. Given
such a decomposition, we choose a parametrization of the co-tetrad which
leads to the usual ADM parametrization in terms of the metric,
$g_{\mu\nu} := e^I_{\mu} e^J_{\nu}\eta_{IJ}$. Both the co-tetrad and the
corresponding tetrad in this parametrization are displayed below
\begin{gather}
e^I_{t}   =   Nn^I + N^aV_a^I  ,\qquad  e^I_{a}  = V^I_{a} , \qquad n^In_I =
-1  , \qquad  n^IV^J_{a} \eta_{IJ} = 0, \label{CoTetrad} \\
e_I^{t}   =   - N^{-1} n_I  , \qquad  e^{a}_I  =  N^{-1} n_I N^{a} + V_I^{a}
 ,\qquad  n^I V_I^a = 0 , \nonumber \\
\mathrm{with} \quad  V^a_I V^J_a  =  \delta_I^J + n_I n^J  , \qquad V^a_I V^I_b
= \delta^a_b . \label{Tetrad}
\end{gather}

We will now restrict to conf\/igurations such that $n_i = 0$, $n_0 = -1$.
This also implies that $V_{a}^0 = 0 = V_0^{a}$ and that $V^a_i$ are
invertible with $V_a^i$ as the inverse\footnote{This would correspond to
the choice of the so-called {\em time gauge} if we started without
restricting the conf\/igurations a priori.}. We also def\/ine the 3-metric
$q_{ab} := V_a^i V_b^j \delta_{ij}$ (which is positive def\/inite in
classical theory) and denote $q := \det(q_{ab})$.

For future convenience we introduce $\Psi := q^{1/4}\lambda$,
$\Psi^{\dagger} := q^{1/4}\lambda^{\dagger}$. This absorbs away the
$\sqrt{q}$ factors in the Lagrangian as well as in the constraints. Note
that the terms involving the derivatives of~$\sqrt{q}$ cancel out. The
$\lambda$ fermionic variables being of density weight zero, the~$\Psi$
fermionic variables are of density weight~1/2. From now on we will use
the half density variables.

Substituting the 3+1 parametrization of the tetrad and using the
time-gauge, the Lagrangian can be written as
\begin{gather*}
{\cal L}_{\mathrm{Dirac}}   =   \frac{i}{2} ( \Psi^{\dagger} \partial_t
\Psi - \partial_t (\Psi^{\dagger}) \Psi) - \omega_{t0i} {\cal G}_F^{0i} -
\frac{1}{2} \omega_{tij} {\cal E}^{ijk}{\cal G}_{k}^F - N^{a'} {\cal
H}^F_{a'} - N {\cal H}_F,
\end{gather*}
where
\begin{gather*}
{\cal G}^{0i}_F   =   0  ,\qquad  {\cal G}^F_i    =
- \frac{i}{2} {\cal E}_{ijk} \Psi^{\dagger}\sigma^{jk} \Psi  =
\Psi^{\dagger}\gamma_5\sigma_{0i}\Psi, 
\\
{\cal H}^F_{a'}   =   - \frac{i}{2}\left( \bar{\Psi}\gamma^0 D_{a'} \Psi
- \overline{D_{a'} \Psi}\gamma^0 \Psi \right) ,\qquad
{\cal H}_F   =   \frac{i}{2} V^a_i \left( \bar{\Psi}\gamma^i D_a \Psi  -
\overline{D_a \Psi}\gamma^i \Psi \right).
\end{gather*}

For the fermions, the action being linear in velocities, we have primary
constraints, $\pi_{\lambda} \sim \lambda^{\dagger}$,
$\pi_{\lambda^{\dagger}} \sim \lambda$, which are second class. Also
these variables fail to be Darboux coordinates~-- do not have vanishing
Poisson brackets\footnote{Strictly {\em Generalized Poisson brackets}
\cite{HennauxTeitelbaum}, due to the Grassmann nature of the fermions.}
with the gravitational variables due to the $\sqrt{q}$ factor. The shift
to $\Psi$, $\Psi^{\dagger}$ variables makes the matter and gravitational
variables Poisson-commute. Def\/ining Dirac brackets relative to these
primary, second class constraints allows us to use $\Psi$,
$\Psi^{\dagger}$ as basic variables with Dirac brackets given by
\begin{gather*}
\big\{ \Psi^{\alpha}(x),  \Psi^{\dagger}_{\beta}(y) \big\}  =
- i\delta^{\alpha}_{\beta} \delta^3(x, y). 
\end{gather*}

From the details given in \cite{DateNew}, we have the f\/inal expressions:
\begin{gather} \label{FinalDefns}
P^a_i  := (\kappa\gamma)^{-1} E^a_i = (\kappa\gamma)^{-1}\sgn
V^a_i\sqrt{q},\qquad  A^i_a = \gamma\,\sgn K_a^i - \Gamma^i_a(V),
\nonumber\\
\Gamma^i_a (V)  :=  \frac{{\cal E}^{ijk}}{2}V_k^b\big\{\partial_b
V_{aj} - \partial_a V_{bj} + V^c_j V^l_a \partial_b V_{cl}\big\},
\nonumber\\
\{ A^i_a(x) , P^b_j(y) \} =  \delta^b_a\delta^i_j \delta^3(x, y)
,\qquad  \{\Psi^{\alpha}(x) , \Psi^{\dagger}_{\beta}(y) \}_+  =  - i
\delta^{\alpha}_{\beta}\delta^3(x,y),
\nonumber\\
{\cal G}_i  =  \partial_a P^a_i + {\cal E}_{ij}^{~k} A^j_a P^a_k -
\frac{i}{2}{\cal E}_{ijk}\Psi^{\dagger}\sigma_{jk}\Psi ,\qquad
\sigma_{jk} := \frac{1}{4}[\gamma_j, \gamma_k],
\nonumber\\
{\cal H}_a =  F^i_{ab}P^b_i - \frac{i}{2}\left(\bar{\Psi}\gamma^0
{\cal D}_a \Psi - \overline{{\cal D}_a\Psi} \gamma^0 \Psi\right) ,\qquad
{\cal D}_a \Psi := \left(\partial_a -i A^i_a \gamma_5\sigma_{0i}\right)
\Psi, \nonumber\\
{\cal H} :=   \kappa\gamma^2 \frac{1}{2}\frac{P^b_j
P^c_k}{\sqrt{q}}\big\{ {\cal E}^{jk}_{~~l}F^l_{bc}({A}) - (1 +
\gamma^2) \big( {K}^j_b {K}^k_c - {K}^j_c {K}^k_b\big) \big\} +
\gamma\partial_a\left(\sgn  V^a_i{\cal G}^i_{\rm vac}\right) \nonumber \\
\phantom{{\cal H} :=}{} + \frac{i}{2}\frac{\kappa\gamma\, \sgn P^a_i}{\sqrt{q}}\left\{
\bar{\Psi}\gamma^i{\cal D}_{a}\Psi - \overline{{\cal
D}_{a}\Psi}\gamma^i\Psi \right\}({A}) + \left[
\left(\frac{\kappa\gamma^2}{2 \sqrt{q}}{K}^i_aP^a_i\right)
\big(\Psi^{\dagger}\gamma_5\Psi\big)\right] \nonumber \\
\phantom{{\cal H} :=}{}
  - \left[
\frac{3}{16}\kappa\frac{\big(\Psi^{\dagger}\gamma_5\Psi\big)^2}{\sqrt{q}}
-~\kappa\frac{(\Psi^{\dagger}\gamma_5\sigma^{0i}\Psi)(\Psi^{\dagger}\gamma_5\sigma^{0}_{\
i}\Psi)}{\sqrt{q}} \right]. \label{FinalHam}
\end{gather}
In the above, ${\cal G}_i^{\rm vac} = \partial_aP^a_i + {\cal
E}_{ij}^{~k}A^j_a P^a_k  := {\cal G}^i - {\cal G}^i_F$.

\begin{remark}
 {\em Dimensionally}, $\kappa \sim L^2$, $(A, K,
\partial) \sim L^{-1}$, $E \sim L^0$, $P \sim L^{-2}$, $(\Psi,
\bar{\Psi}) \sim L^{-3/2}$, ${\cal G} \sim L^{-3}$, $({\cal H}_a, {\cal
H}) \sim L^{-4}$.

The {\em density weights are:} $( P, {\cal G}, {\cal H}_a, {\cal H} ) =
+1$, $(\Psi, \bar{\Psi}) = 1/2$ and $(A, K, V, \Gamma) = 0$.

Under\footnote{Discussed in Subsection \ref{Parities}.} {\em Lorentz
parity:} $(V, K, \sgn)$ are odd, $(\Gamma, A, P, \Psi, \bar{\Psi})$ and
the constraints are all even.

Under {\em parity combined with $\gamma \to - \gamma$:} $(V, K, \Gamma,
A, P, \gamma\,\sgn)$ and the constraints are all even.

The $A$ and $K$ above correspond to the vacuum case for which the
Thiemann identities hold.  The inverse square root of $q$ and $K_a^i$
appearing above are manipulated exactly as in the vacuum case.

Explicitly, the identities are
\begin{gather*}
\sgn {\cal E}^{bca} V^i_a  =  {\cal E}^{ijk}\frac{E^b_j
E_k^c}{\sqrt{\det (E^a_i)}}   , \qquad q
:= (\det(V^i_a))^2 = \det(E^a_i),
\\
\kappa\gamma\frac{\sgn}{2}V^i_a(x)   =   \left\{{A}^i_a(x), \int d^3y
\sqrt{q}\right\}  \quad \Rightarrow    \nonumber \\
{\cal E}^{ijk}\frac{E^b_j E_k^c}{\sqrt{\det(E^a_i)}}  =
\frac{2}{\kappa\gamma} {\cal E}^{bca} \left\{{A}^i_a(x), \int d^3y
\sqrt{q}\right\},
\\
H_E(1)   :=   \frac{\kappa\gamma^2}{2}\int \frac{P^b_j
P^c_k}{\sqrt{q}}{\cal E}^{jk}_{~l}F^l_{bc}   , \qquad  \overline{{K}}   :=
\int d^3y \, \sgn {K}^i_a P^a_i \quad \Rightarrow \\
\overline{{K}}  =  (\kappa\gamma^3)^{-1}\left\{ H_E(1), \int d^3y
\sqrt{q} \right\},\qquad
\sgn {K}^i_a(x)  =  \{{A}^i_a(x), {\overline{K}}\}.
\end{gather*}
These identities suf\/f\/ice to derive a quantization the Hamiltonian
constraint from that of the `Euclidean Hamiltonian constraint' (the
f\/irst term in the Hamiltonian constraint) and of the volume operator.
This completes the classical canonical formulations as it follows from
the action.
\end{remark}

\subsection{Constraint algebra}

It is easy to see that the gauge constraint generates correct gauge
transformation of the basic f\/ields. Specif\/ically, with ${\cal
G}(\Lambda) := \int_{\Sigma_3}d^3x \Lambda^i {\cal G}_i$,
\begin{gather*}
\{ A^i_a(x), {\cal G}(\Lambda)\}  =  - {\cal D}_a \Lambda^i  =  -
\partial_a \Lambda^i - {\cal E}^i_{~jk}A^j_a\Lambda^k, \\
\{ P_i^a(x), {\cal G}(\Lambda)\}  =  + {\cal E}_{ij}^{~k}\Lambda^j
P^a_k, \\
\{ \Psi^{\alpha}(x), {\cal G}(\Lambda)\}  =  -i \Lambda^i
(\gamma_5\sigma_{0i}\Psi)^{\alpha}, \\
\{ \Psi^{\dagger}_{\alpha}(x), {\cal G}(\Lambda)\}  =  +i \Lambda^i
(\Psi^{\dagger}\gamma_5\sigma_{0i})_{\alpha}.
\end{gather*}

If we compute the inf\/initesimal action of the ${\cal H}_a$ constraint on
the basic variables, we see that it equals the Lie derivatives of the
basic variables only up to an SU(2) gauge transformation.  We are
however free to modify the constraints by adding suitable combinations
of themselves.  So we {\em define} the dif\/feomorphism constraint as
\begin{gather*}
{\cal C}(\vec{N})  :=  \int_{\Sigma_3}d^3x N^a {\cal C}_a \quad
\mathrm{with} \nonumber \\
{\cal C}_a  :=  {\cal H}_a - A^i_a {\cal G}_i  =  P^b_i\partial_a
A^i_b - \partial_b(A^i_a P^b_i) +
\frac{i}{2}\big(\Psi^{\dagger}\partial_a \Psi -
\partial_a\Psi^{\dagger}\cdot\Psi\big),
\end{gather*}
which leads to the inf\/initesimal transformations,
\begin{gather*}
\{ A^i_a(x), {\cal C}(\vec{N})\}  =  {\cal L}_{\vec{N}} A^i_a  =
\partial_a(N^bA^i_b) + N^b(\partial_b A^i_a - \partial_a A^i_b), \\
\{ P_i^a(x), {\cal C}(\vec{N})\}  =  {\cal L}_{\vec{N}} P_i^a =
N^b\partial_b P^a_i - P_i^b \partial_b N^a + 1\cdot(\partial_b N^b)
P^a_i, \\
\{ \Psi^{\alpha}(x), {\cal C}(\vec{N})\}  =  {\cal L}_{\vec{N}}
\Psi^{\alpha}  =  N^b\partial_b \Psi^{\alpha} +
\frac{1}{2}\cdot(\partial_b N^b)\Psi^{\alpha}, \\
\{ \Psi^{\dagger}_{\alpha}(x), {\cal C}(\vec{N})\} =  {\cal
L}_{\vec{N}} \Psi^{\dagger}_{\alpha}  =  N^b\partial_b
\Psi^{\dagger}_{\alpha} + \frac{1}{2}\cdot(\partial_b
N^b)\Psi^{\dagger}_{\alpha}.
\end{gather*}

This implies that $\{{\rm var}, \int N^a C_a\} = {\cal L}_{N^a}({\rm var})$ for all
variables. The Gauge constraint already generates the correct gauge
transformation of the basic variables. By inspection, it follows that
the gauge constraints (weakly) commute with the dif\/feomorphism and the
Hamiltonian constraint, the gauge constraint and the dif\/feomorphism
constraints form sub-algebras and the dif\/feomorphism constraint
transforms the Hamiltonian constraint by the Lie derivative. The
non-trivial bracket is the bracket of two Hamiltonian constraints.

Before turning to loop quantization, we brief\/ly draw attention to the
invariance (or lack of it) under two distinct `parity' operations.

\subsection{Parity and Lorentz parity}\label{Parities}
Recall that we begin with the (co-)tetrad f\/ield $e^I_{\mu}$, the Lorentz
connection $\omega^{IJ}_{\mu}$ and the fermion f\/ields $\lambda$,
$\bar{\lambda}$ (or $\Psi$, $\bar{\Psi}$) def\/ined over a manifold $M \sim
\mathbb{R}\times\Sigma_3$ which is assumed to be orientable. With the
topology specif\/ied, $M$ can be taken to be {\em time-orientable} with
respect to all the metric tensors constructed by the parametrization
(\ref{CoTetrad}), (\ref{Tetrad}).  Obviously, $\Sigma_3$ is orientable as
well.

There are two distinct sets of `parity' transformations: orientation
reversing dif\/feomorphism of $M$ and a ${\rm O}(1,3)$ transformation with
determinant $= -1$.  Note that these distinctions do not exist in f\/lat
space-time with Minkowski metric where the allowed dif\/feomorphisms are
the isometries of the Minkowski metric which are the same as ${\rm O}(1,3)$
transformations.  We will keep the time orientation f\/ixed. Orientation
reversing dif\/feomorphisms of~$M$ will then be reversing the orientation
of~$\Sigma_3$. We will refer to these as {\em parity} transformations.
The improper Lorentz transformations~$\Lambda^I_J$, will also be taken
so that $\det \Lambda = -1$ and $\Lambda^0_{~0} = 1$ and will be
referred to as {\em Lorentz parity transformations}.

After going to the canonical framework in the `time gauge', we have the
f\/ields $A^i_a$, $P^a_i$, $\Psi$, $\bar{\Psi}$ def\/ined on $\Sigma_3$. A parity
transformation is now an orientation reversing dif\/feomorphism of~$\Sigma_3$ while the improper ${\rm O}(3)$ transformation, {\em inversion},
will be taken as the Lorentz parity transformation.

In the Lagrangian framework, the Hilbert--Palatini action is invariant
under both sets of transformations while the Nieh--Yan (and the
Pontryagin) actions {\em change signs under parity} but are {\em
invariant under Lorentz parity.} Hence the combined action is invariant
under Lorentz parity and {\em non}-invariant under parity. The $\sgn$
factor in the Hilbert--Palatini action is crucial for this.

The variables of the canonical framework are def\/ined in terms of those
of the Lagrangian framework. These def\/initions of the ${\rm SU}(2)$ connection
in terms of $K$ and $\Gamma$ and the conjugate momentum in terms of the
triad {\em are} consistent with the ${\rm SO}(3)$ gauge transformation
extended to include the Lorentz parity. Thus the triad which transforms
by the def\/ining representation changes sign under Lorentz parity. The
`densitised triad' (or the conjugate momentum) transforms by the adjoint
representation and should be invariant under Lorentz parity. The $\sgn$
factor in their def\/initions precisely takes care of this. The same can
be seen in the def\/inition of the connection. It is easy to see that the
symplectic structure and the constraints (vacuum) are {\em all invariant
under Lorentz parity}.

When fermions are included, these are {\em scalars} under orientation
preserving dif\/feomorphisms\footnote{Def\/inition of spinors depends on the
orientation. Thus the action of parity on spinors needs a much more
careful statement. We will assume for simplicity that the under
orientation reversing dif\/feomorphism, the fermionic action remains
invariant.}, and transform as $\Psi \to \gamma^0 \Psi$ under Lorentz
parity. All the constraints including fermions are invariant under
Lorentz parity. This is true in both the Lagrangian and the canonical
frameworks.

With regards to parity the situation is dif\/ferent. The action is not
invariant under parity, due to the Nieh--Yan term. In the canonical
framework, the connection is not simply even/odd under parity since the
$K$ term changes sign while the $\Gamma$ does not. The `densitised
triad' also acquires an extra minus sign under parity (behaves as a
`pseudo-vector of weight 1'). The symplectic structure thus is {\em not}
invariant. The constraints also are not invariant under parity and this
is consistent with the non-invariance of the action.

The action {\em is} invariant under parity {\em combined with $\gamma
\to - \gamma$}. Our def\/initions have the appropriate factors of $\gamma$
to restore the simple (even) behaviour of the basic canonical variable
resulting also in the invariance of the Poisson brackets and
constraints.

In short, {\em Lorentz parity is an invariance of the action as well as
the canonical framework while parity is not. However, parity combined
with $\gamma \to - \gamma$ is an invariance of both action and the
canonical framework.}

\begin{remark} We could consider a canonical formulation {\em ab
initio}, say by making a canonical transformation from the ADM
variables, without any reference to an action.  We could then {\em
define} of the basic canonical variables $A^i_a$, $E^a_i$, without the
factors of $\sgn$.  This will restore the `densitized triad' to its
usual density weight~1 vector density status and the connection to its
1-form status.  The canonical framework is then {\em invariant} under
parity (without changing sign of~$\gamma$).  Under Lorentz parity,
however, the connection does not have simple behaviour and the conjugate
momentum is odd. The reason of course is that the internal index `$i$'
on the triad and on the extrinsic curvature refers to the def\/ining
representation while that on the connection and its conjugate refer to
the adjoint representation and these are distinguished by the improper
orthogonal transformations. This results in non-invariance of the
canonical framework under Lorentz parity. If the sign of $\gamma$ is
changed along with the Lorentz parity transformation, then the basic
variables are even, the symplectic structure is invariant and so are the
constraints {\em and the action}. Thus, the def\/initions without the
$\sgn$ factors, interchanges the role of Lorentz parity and parity
appropriately combined with $\gamma \to - \gamma$.

Which of these notions is `appropriate'?  If we were to consider
formulation in terms of the metric tensor, then the notion of Lorentz
parity is not even def\/inable as there is no internal Lorentz
transformation.  On the other hand, with fermionic matter we {\em have
to} consider oriented manifolds and the tetrad formulation, introducing
the possibility of Lorentz parity. If the orientation of the manifold is
regarded as a f\/ixed background structure, the parity transformations are
excluded by def\/inition and Lorentz parity alone remains.  Which of these
notions is {\em relevant} for experimental observation, is unclear and
so is the issue of `gravitational parity violation/invariance' in LQG.
\end{remark}

In the next subsection, we brief\/ly summarise the loop quantization of
fermions \cite{BojowaldDas,BojowaldDas2,ThiemannScalar,ThiemannQSD,ThiemannQSD2,ThiemannQSD4}.

\subsection{Loop quantization of fermions}\label{LoopQuantization}

The loop quantization of the gravitational f\/ields as well as of the
fermions has been already given by Thiemann and those arguments remain
valid. We have already gone from $\lambda$ fermions to the half density
$\Psi$ fermions. In our derivation from the action, this is simply seen
as the choice of Darboux coordinates. We also do not work with the
`conjugate' variables $\pi_{\Psi}$, $\pi_{\bar{\Psi}}$. Instead we solve
the primary second class constraints and use Dirac brackets relative to
these.  Thus we take, for a single Dirac fermion, $\Psi^A$, $\bar{\Psi}_A$
to be complex, Grassmann valued, half densities: $\bar{\Psi}_A :=
(\Psi^A)^*$, $A = 1, \ldots, 4$. This can easily be generalized to many
Dirac fermions or Weyl fermions\footnote{The overbar on a $\Psi$ with an
index $A$, denotes the complex conjugate while one without an index
denotes the Dirac conjugate: $\bar{\Psi} = \Psi^{\dagger}\gamma^0$.}.

As mentioned in the introduction, the f\/irst step is to construct
functions on the fermionic phase space.  Here we discuss Thiemann's
proposal presented in a slightly dif\/ferent form. It is based on a
particular, naturally available, class of smearing densities.

Consider a decomposition of $\Sigma_3$ into cells $B_n$, closed subsets
of $\Sigma_3$. Let $\epsilon^3 := \mu(B_n)$ be the {\em Lebesgue
measure} (`coordinate volume') of the cell $B_n$ and let $v_n$ be the
{\em centre} of $B_n$ (a marked point in the interior of $B_n$).  Def\/ine
the {\em indicator} or ({\em characteristic}) function:
\begin{gather*}
\chi_{\epsilon}(v_n, x) :=  \chi_n(x) :=   \begin{cases} 1, &
\mathrm{if} ~ x \in \mathrm{Int}(B_n), \\ 0, & \mathrm{otherwise}.
\end{cases}
\end{gather*}
Then the quantity, $\chi_n(x)/\epsilon_n^{3w}$ has density weight~$w$.
This is because, the Lebesgue measure of a cell, transforms as a scalar
density of weight~$-1$ under a coordinate transformation and the indicator
function is of course invariant. Note that eventually we are interested
in the limit of inf\/inite ref\/inement of the cell decomposition,
$\epsilon_n \to 0$, and the stated behaviour of the Lebesgue measure is
valid in this limit.  Using these quantities, labelled by the cells of a
cell decomposition, we can smear any scalar density of weight $(1 - w)$
and construct a function on the conf\/iguration/phase space.

For a Grassmann variable $\Psi$ (the spinorial index is suppressed) of
weight~1/2, we def\/ine the Grassmann valued functions
\begin{gather*}
\theta_n (\Psi)   :=   \int_{\Sigma_3} d^3x\,
\frac{\chi_n(x)}{\epsilon_n^{3/2}} \Psi(x)   =   \int_{B_{n}} d^3x\,
\frac{\chi_n(x)}{\epsilon_n^{3/2}} \Psi(x).
\end{gather*}

Under a dif\/feomorphism: $B_n \to B'_n$, its Lebesgue measure
$\epsilon^3_n \to \epsilon^{'3}_n = |\Case{\partial x'}{\partial x}|
\epsilon_n^3$, $d^3x' = |\Case{\partial x'}{\partial x}| d^3x$ while
$\Psi'(x') = \sqrt{|\Case{\partial x}{\partial x'}|} \Psi(x)$. This
implies that
\begin{gather*}
\theta_n'  :=   \int_{B'_n}d^3x' \frac{\chi_n(x')
\Psi'(x')}{\epsilon_n^{'3/2}}   =   \int_{B_n}d^3 x
\frac{\chi_n(x)\Psi(x)|\Case{\partial x'}{\partial x}||{\Case{\partial
x}{\partial x'}}|^{1/2}}{\epsilon_n^{3/2}|\Case{\partial x'}{\partial
x}|^{1/2}} \nonumber \\
\phantom{\theta_n'}{}  =  \int_{B_n} d^3x \frac{\chi_n(x) \Psi(x)}{\epsilon^{3/2}_n}   =
\theta_n.
\end{gather*}
Thus, the $\theta$ variables are indeed {\em invariant} under
dif\/feomorphism\footnote{In the units where $\hbar = 1 = c$, these
variables are also {\em dimensionless}. This is because the action is
dimensionless which implies $\Psi \sim L^{-3/2}$ and $\epsilon \sim L$.
This is similar to the holonomy variables constructed from connections.
For the similar variables corresponding to scalar f\/ields, this is not
so.}. It is immediate that $\bar{\theta}_n = (\theta_n)^*$.

In the limit of f\/iner cell decomposition, $\epsilon_n \to 0$, we get the
following expressions:
\begin{gather*}
\lim_{\epsilon_n \to 0} \chi_{\epsilon_n}(v_n, x)   =   \delta_{v_n,x}
\qquad (\mathrm{Kroneker~delta}), \\
\lim_{\epsilon_n \to 0} \frac{\chi_{\epsilon_n}(v_n, x)}{\epsilon_n^3}
:=   \delta^3(v_n, x) \qquad (\delta\text{-distribution}).
\end{gather*}
From these it follows that
\begin{gather*}
\{\theta_m,  \bar{\theta}_n\}   =   \int_{B_m}d^3x\int_{B_n}d^3y
\frac{\chi_m(x)}{\epsilon_m^{3/2}} \frac{\chi_n(y)}{\epsilon_n^{3/2}} \{
\Psi(x), \bar{\Psi}(y) \} \nonumber \\
\phantom{\{\theta_m,  \bar{\theta}_n\}}{}  =   -i \int_{B_m}d^3x\int_{B_n}d^3y \frac{\chi_m(x)}{\epsilon_m^{3/2}}
\frac{\chi_n(y)}{\epsilon_n^{3/2}} \delta^3(x,y)  \nonumber \\
\phantom{\{\theta_m,  \bar{\theta}_n\}}{}  =   - i \delta_{m,n} \int_{B_n}d^3x \frac{\chi_n(x)}{\epsilon_n^3}
 =   -i \delta_{m,n} , \qquad \because \chi_m(x) \chi_n(x) = \delta_{mn}\chi_m(x) .
\end{gather*}
With these, we obtain the Poisson brackets among the $\theta$ variables
(functions on the phase space). The identity in the third line above
holds because the $\chi_n(x) = 0$ unless $x$ is in the {\em interior} of
the cell $B_n$.

At f\/inite but small $\epsilon_n$, we can write,
$\theta_n   \approx   \epsilon_n^{3/2}\Psi(v_n)$. Notice that the
$\theta_n$ variables are ef\/fectively associated with the points $v_n$.
This is particularly useful in expressing the constraint expressions in
terms of the $\theta$'s as illustrated below.

Recall from equation (\ref{FinalHam}) that the smeared Hamiltonian
constraint has fermion bilinears with and without derivative as well as
terms quartic in fermions. Using a cell decomposition, we can express
the integral as a sum by restricting the integrals over the cells. In
the limit of small~$\epsilon_n$, these integrals can be approximated by
using the mean value theorem. The factors of~$\epsilon_n$'s available
can be distributed with the fermions to go from~$\Psi(v_n)$ to~$\theta_n$ variables. The correct density weight of +1 ensures that no
factors of~$\epsilon_n$'s remain unabsorbed. In equations
\begin{gather*}
\int_{\Sigma_3}d^3x \bar{\Psi}_A(x) M^A_{~B}(x)\Psi^B(x)  \approx
\sum_n\epsilon_n^3\bar{\Psi}_A(v_n)M^{A}_{~B}(v_n) \Psi^B(v_n) \nonumber
\\
\qquad{}  =   \sum_n\big(\epsilon_n^{3/2}
\bar{\Psi}_A(v_n)\big)M^{A}_{~B}(v_n)
\big(\epsilon_n^{3/2}\Psi^B(v_n)\big)
  \approx   \sum_n\bar{\theta}_A(v_n)M^{A}_{~B}(v_n) \theta^B(v_n), \\
\int_{\Sigma_3}d^3x \bar{\Psi}_A(x) (\sigma^i
f^a_i(x))^A_{~B}\frac{\partial\Psi^B(x)}{\partial x^a}    \approx
\sum_n\epsilon_n^3\bar{\Psi}_A(v_n)(\sigma^i f^a_i(v_n))^{A}_{~B}
\frac{\partial \Psi^B(v_n)}{\partial v^a_n} \nonumber \\
\qquad{}  \approx   \sum_n\bar{\theta}_A(v_n)(\sigma^i f^a_i(v_n))^{A}_{~B}
\frac{\partial \theta^B(v_n)}{\partial v^a_n}.
\end{gather*}
This takes care of the fermionic bilinears.  The quartic terms have a
factor of $\sqrt{q}(x)$ in the denominator.  Following
\cite{ALReview,BojowaldDas,BojowaldDas2}, we f\/irst express it as a Poisson bracket
with only positive powers of the volume of the cell containing the point~$x$. From the def\/inition of the determinant and our conventions given in
the footnote~\ref{Notation}, it follows that
\[
\frac{\sgn}{\sqrt{q}}   =   \frac{1}{6} {\cal E}^{abc}{\cal E}_{ijk}
\frac{V^i_a V^j_b V^k_c}{q}.
\]
Let ${\cal V}_n := \int_{B_n}d^3x \sqrt{q}(x)$. Then
\begin{gather*}
\{ A^i_a(x), {\cal V}_n^{1/3}\}   =   \frac{1}{3} {\cal V}_n^{-2/3}\{
A^i_a(x), {\cal V}_n\}   = \begin{cases}
 \dfrac{\kappa\gamma}{6} V^i_a {\cal
V}_n^{-2/3}, & \mbox{if $x \in B_n$},\\
0, &  \mbox{otherwise },
\end{cases} \nonumber \\
\therefore \frac{\sgn}{\sqrt{q}(x)}   =
\frac{36}{(\kappa\gamma)^3}\{A^i_a, {\cal V}_n^{1/3}\}\{A^j_b, {\cal
V}_n^{1/3}\}\{A^k_c, {\cal V}_n^{1/3}\} {\cal E}^{abc}{\cal
E}_{ijk}\frac{{\cal V}^2_n}{q(x)}.
\end{gather*}
For small $\epsilon_n$, and for non-zero answer ($x \in B_n$), ${\cal
V}_n \approx \epsilon_n^3 \sqrt{q}(x)$ and therefore ${\cal V}^2_n/q(x)
\approx \epsilon^6_n$. These factors of $\epsilon_n$ are neatly combined
with the 4 $\Psi$, $\bar{\Psi}$ variables to go over to the $\theta$
variables.  The $\epsilon^3_n$ from the integration measure combines
with the three $A^i_a$ to produce the combinations ${\rm Tr}({\cal
E}^{IJK}h^{-1}_I\{h_I,\dots\}h^{-1}_J\{h_J, \dots\}h^{-1}_K\{h_K,
\dots\}$) with the ${\cal E}$ ensuring that the three edges are
non-coplanar and the Tr ensuring the SU(2) invariance.

Thus all integral expression in the constraints are expressible in terms
of the $\theta$ variables. Since the transition from the $\Psi$
variables to the $\theta$ variables is a linear operation, the Grassmann
properties of the $\theta$ variables are the same as those of the~$\Psi$
variables. We have thus the classical phase space for Grassmann valued
f\/ields.

To construct a quantum theory, we f\/irst construct the space of complex
valued (and Grassmann valued) functions and def\/ine an inner product on
it. The elementary $\theta$ variables, are to represented as
multiplicative operators while the their complex conjugates (``momenta'')
are to be represented by derivative operators. The procedure is standard~\cite{Berezin,Berezin2,ThiemannScalar,ThiemannQSD,ThiemannQSD2,ThiemannQSD4}. The Grassmann nature restricts the functions
to be polynomials eg for a single pair of complex Grassmann variables,
$\theta$, $\bar{\theta}$, the most general function is: $f(\bar{\theta},
{\theta}) = a + b\theta + c\bar{\theta} + d\bar{\theta}{\theta}$ where
$a$, $b$, $c$, $d$ are complex numbers. A {\em holomorphic} function, has no
$\bar{\theta}$ dependence. For $k$ number of complex pairs, the vector
space of holomorphic functions is $2^k$-dimensional. On this space one
has the usual {\em Berezin} measure, $\int d\bar{\theta}d\theta
f(\bar{\theta},{\theta}) = d$, however as pointed out by Thiemann, this
is not positive def\/inite and thus unsuitable for constructing a Hilbert
space. The modif\/ication proposed by Thiemann is: $\mu(\bar{\theta},
\theta) := e^{\bar{\theta}\theta}d\bar{\theta}d\theta$. The inner
product def\/ined by $\langle f, g\rangle := \int\mu(\bar{\theta},
\theta)f^*(\bar{\theta})g(\theta)$, turns the function space into a~Hilbert space.

On this, def\/ine the operators
\begin{gather*}
[\hat{\theta}f](\theta)  :=  \theta f(\theta) ,\qquad
[\hat{\bar{\theta}}f](\theta)   :=   \hbar\frac{d f(\theta)}{d\theta}.
\end{gather*}
These operators satisfy: $[\hat{\theta}, \hat{\bar{\theta}}]_+ = \hbar$
as desired by the general quantization rule, $\{A, B\} \to
(-i/\hbar)[\hat{A}, \hat{B}]$, valid for the generalized Poisson
bracket. It follows that $\hat{\bar{\theta}} = (\hat{\theta})^{\dagger}
$ also holds ref\/lecting the classical relation $\bar{\theta} = \theta^*$
and $\mu(\bar{\theta}, \theta)$ is the unique normalised measure
selected by the above adjointness property~\cite{ThiemannScalar,ThiemannQSD,ThiemannQSD2,ThiemannQSD4}.
Generalization to f\/initely many Grassmann variables is immediate. For
fermionic f\/ields (inf\/initely many variables) one proceeds via projective
limit.

For $d$-Weyl fermion f\/ields, we will have $2d$ $\theta$ variables,
$\theta^1, \theta^2, \ldots, \theta^{2d}$ at each point of $\Sigma_3$.
The holomorphic functions are polynomials of maximal degree $2^{2d}$ in
the $\theta$ variables at {\em each} point.  Denote a {\em monomial} of
degree $k$ at a point~$v$ by
\begin{gather*}
F_{v, \{i\}_{k(v)}} :=
\theta^{i_1}(v)\theta^{i_2}(v)\cdots\theta^{i_{k(v)}}(v), \qquad i_1,
\ldots, i_k \in [1, 2d], \qquad k(v) \in [0, 2^{2d}].
\end{gather*}

For every f\/inite set of points\footnote{When matter and gravity are
considered together, it is natural (suf\/f\/icient) to take the set of
points to be the {\em vertices} of graph labelling a spin-network
function in the gravity sector (and matter gauge sector if present).},
$\{\vec{v}\} = \{v_1, v_2, \ldots, v_n\}$ and a corresponding vector of
labels, $\vec{I}_{\vec{k}} := \{\{i\}_{k(v_1)}, \{i\}_{k(v_2)}, \ldots,
\{i\}_{k(v_n)} \}$, def\/ine the {\em elementary functions of $\theta$'s},
\begin{gather*}
{\cal F}_{{\vec I}_{\vec k}}   :=   \prod_{j = 1, n} F_{v_j,
\{i\}_{k(v_j)}}.
\end{gather*}
Finite linear combinations of these elementary functions are called
(fermionic) cylindrical functions which are orthonormal with respect to
the inner product def\/ined point-wise. Closure of the set of these
cylindrical functions def\/ines the {\em Hilbert space of the loop
quantization of fermions}. Quantization of more general observables
proceeds by f\/irst expressing these observables in terms of the $\theta$
variables and promoting them to the corresponding operators.

\subsection{Implications for loop quantum cosmology}

Implications of fermionic matter has also been discussed in the context
of homogeneous models by Bojowald and Das in \cite{BojowaldDas,BojowaldDas2}. The
symmetric fermionic f\/ields are only restricted to be constants on the
spatial manifold by homogeneity. Isotropy however requires these
constants to be zero. One way to see this is to note that from the
presumed non-zero $\Psi$'s we can construct spatial vectors as
$V^a_i\bar{\Psi}\gamma^i\Psi$ and by isotropy there cannot be any
non-zero constant vector. For homogeneous, anisotropic models however
non-zero fermionic constants are allowed\footnote{Only fermionic matter
is sensitive to the action of the gravitational gauge group (subgroup of
local Lorentz transformations).  All other matter f\/ields are {\em
scalars} with respect to this action and thus do not contribute to the
Gauss constraint.\label{FootNote1}}.

This has a major implication for diagonal models. Recall that the
homogeneous connection and triad are of the form: $A^i_a(x) =
\Phi^i_I\omega^I_a(x)$, $E^a_i(x) = \sqrt{g_0} P^a_I X^a_I(x)$ where,
$\omega^I_a$ are the Maurer--Cartan forms and $X^a_I$ their dual vector
f\/ields. The sub-class of {\em diagonal} models is def\/ined by the
restriction: $\Phi^i_I := c_I\Lambda^i_I$, $P^I_i := p^I\Lambda^I_i$ (no
sum over $I$), where $\Lambda_i^I \Lambda^i_J = \delta^I_J$,
$\Lambda^i_I\Lambda^j_J\Lambda^k_K{\cal E}_{ijk} = {\cal E}_{IJK}$.  With
this choice, the homogeneous, gravitational part of the Gauss Law
constraint vanishes identically. With the possibility of non-zero,
homogeneous fermionic degrees of freedom, the homogeneous ${\cal G}^i_F
\neq 0$ and therefore the gravitational part {\em cannot} be zero.
Consequently, the connection and the triad variables cannot be
diagonalised simultaneously.

A way out, suggested in~\cite{BojowaldDas,BojowaldDas2} is to use dif\/ferent
$\Lambda$'s for connection and triad. The corresponding dynamical system
is still complicated and a specif\/ic choice is made to get a simplif\/ied
model, wherein the constant fermionic vector is taken to be along,
1-axis. This permits a identif\/ication of canonical coordinates, 4 in
number. To explore parity invariance more explicitly, the model is
further specialized to Bianchi~I LRS class. The main conclusion drawn is
that in this specif\/ic case, parity invariance at the level of underlying
dif\/ference equation is violated only if there are parity violating
interactions in the {\em matter sector}. However, the dif\/ference
equation remains deterministic even when parity is violated~\cite{BojowaldDas,BojowaldDas2}.

\section{Gauge f\/ields} \label{YM}

The construction of the kinematical Hilbert space for gauge f\/ields
proceeds exactly as in the case of the gravitational, SU(2) gauge
f\/ields. For identical logic, in a background independent context, the
gauge f\/ields are quantized through the use of the holonomies and f\/luxes
of the electric f\/ields. The new feature is the structure of the
corresponding contributions to the Hamiltonian constraint and of course
the new Gauss constraints corresponding to the additional gauge
invariances. As remarked in the footnote~\ref{FootNote1}, there is no
contribution to the rotation constraint.  The contribution to the
dif\/feomorphism constraint is of the same form as in the gravitational
sector. The couplings of these matter f\/ields to the gravitational ones
are through the triad variables for minimally coupled matter and there
are also non-gravitational couplings among the matter f\/ields. The
underlying background independence which forces the use of holonomy and
f\/luxes impacts the quantization of these interaction terms as well. Loop
quantization of gauge f\/ields has been discussed by Thiemann~\cite{ThiemannScalar,ThiemannQSD,ThiemannQSD2,ThiemannQSD4} and Ashtekar--Lewandowski~\cite{ALReview} (see also~\cite{Alfaro}).  The salient points are noted below.

\subsection{Maxwell f\/ield theory}
The gauge group in the Maxwell case case is Abelian which simplif\/ies
some of the technical details. In particular, since the holonomies are
simple numbers (not matrices), the Poisson bracket of holonomy with
3-dimensionally smeared electric f\/ield is proportional to the holonomy
and the 3-smeared electric f\/ield maps cylindrical functions to
cylindrical functions~\cite{ALReview}. We will however use 2-dimensional
smearing and the f\/luxes Poisson commute in both cases.

The classical analysis beginning with the Maxwell action followed by
canonical formulation using the time gauge parametrization is
summarised below
\begin{gather*}
{\cal L}_{\mathrm {Maxwell}}  :=  \frac{1}{16\pi}\sqrt{|g|}
g^{\mu\alpha}g^{\nu\beta}\mathbb{F}_{\mu\nu}\mathbb{F}_{\alpha\beta}, \\
H(N, N^a, \Lambda)  =  \int
\frac{N}{8\pi}\frac{q_{ab}}{\sqrt{q}}\big(\mathbb{P}^a\mathbb{P}^b +
\mathbb{B}^a\mathbb{B}^b\big) +
\frac{N^a}{8\pi}\mathbb{F}_{ab}\mathbb{P}^b +
\frac{\Lambda}{8\pi}\partial_a\mathbb{P}^a,  \\
\mathbb{B}^a  :=  \frac{1}{2}{\cal E}^{abc}\mathbb{F}_{bc},\qquad
\mathbb{E}^a := \mathbb{P}^a   =   - \sqrt{q}\ N\ \mathbb{F}^{ta}
\nonumber
\end{gather*}
with $\mathbb{A}_a$, $\mathbb{P}^b$ being the canonical coordinates.
Note that $\mathbb{P}^a$, $\mathbb{B}^a$ are both of density weight~1.  As
usual, before quantization, we need to express the constraints (and
other observables of interest), in terms of the holonomies and f\/luxes.

We have the common factor of $q_{ab}/\sqrt{q}$. As before, introduce a
cell decomposition with cells~$B_n$, containing points $v_n$ and with
Lebesgue measure $\epsilon_n^3$ and def\/ine ${\cal V}_n := \int_{B_n}
d^3x \sqrt{q(x)}$. Noting that
\begin{gather*}
\{A^i_a(x), {\cal V}_n^l\} = l {\cal V}_n^{l - 1}
\frac{\kappa\gamma}{2}\sgn V^i_a  ,\quad l \in (0, 1)  ,\qquad  {\cal V}_n
\approx \epsilon_n^3\sqrt{q}(v_n) \quad \mathrm{for~small} \ \ \epsilon_n,
\end{gather*}
we write
\begin{gather*}
\frac{q_{ab}}{\sqrt{q}}(x)  \approx
\frac{16}{(\kappa\gamma)^2}\epsilon_n^3  \delta_{ij}\big\{A_a^i(x),
\sqrt{{\cal V}_n}\big\} \big\{A_b^j(x), \sqrt{{\cal V}_n}\big\}.
\end{gather*}

Next, observe that for 2-smeared f\/luxes of $\mathbb{P}^a$, we have
$\mathbb{\Phi}_S := \int_S \mathbb{P}^a {\cal E}_{abc} dS^{bc} \approx
\mathbb{P}^a(v) n_a(S) \epsilon_S^2$ where $S$ is a (small) surface, $v$
is a point in it and $\epsilon^2_S$ (small) is its Lebesgue measure. The
`normal' to $S$ is given by, $n_a(S) := {\cal E}_{abc}\Case{\partial
x^b}{\partial \xi_{\alpha}}\Case{\partial x^c}{\partial
\xi_{\beta}}{\cal E}^{\alpha\beta}$ where $\xi^{\alpha}$, $\alpha = 1, 2$
denote the local coordinates {\em on} $S$ while $x^a(\xi)$ denote the
embedding of the 2-surface in $\Sigma_3$. For a small triangular
surface, we can also replace $n_a(S)\epsilon_S^2 \approx
\Case{1}{2}{\cal E}_{abc}\delta S^b_I\delta S^c_J$ where $\delta S^b_I$,
$\delta S^c_J$ denote coordinate lengths of a pair of edges of the
triangle. Likewise, for an inf\/initesimal curve $e$, $h_e :=
\mathrm{Pexp}(\int_e A) = \exp{(\int_e A)}$ implies $h^{-1}_e \{
h_e, \dots \} \approx \{\int_e A, \dots \} \approx \{\delta t
\dot{e}^aA^i_a\tau_i , \dots \}$.

\looseness=1
To regularise the Hamiltonian, we introduce a cell decomposition and
approximate the integral by a sum-over-cells. This provides a
$\epsilon^3$ and the $q_{ab}/\sqrt{q}$ expression provides another
factor of $\epsilon^3$ for each term. The 6 factors of $\epsilon$ can be
distributed as 2 each for the two $\mathbb{P}$'s and 1 each for the two
gravitational connections in the Poisson brackets. To express the
variables in terms of f\/luxes and holonomies, we need to choose two
surfaces and two curves paying attention to the {\em contraction} of $a,
b$ indices which is part of specif\/ication of gravitational coupling.

We recall from the construction done in the gravitational sector. We
choose a triangulation of $\Sigma_3$ by elementary tetrahedra $\Delta_n$
\cite{ThiemannScalar,ThiemannQSD,ThiemannQSD2,ThiemannQSD4}. In anticipation of the quantization step, we choose
the edges of the tetrahedra to be analytic. {\em One} vertex of each
tetrahedron is distinguished and at the most f\/initely many tetrahedra
meet at such a  vertex.  Edges meeting at a vertex are taken to be
out-going. The three edges of each tetrahedron, provide three linearly
independent tangent vectors at its vertex $v_n$, $\dot{e}^a_{~I}$, $I  =
1, 2, 3$, such that the $a^{\rm th}$ coordinate interval of the $I^{\rm th}$
edge $ = \delta t_I\dot{e}^a_I$, no sum over $I$ and $I$ is {\em not} to
be confused with the Lorentz index which is no longer relevant now.
$\delta t_I$ refers parametrization of the $I^{\rm th}$ edge. Each
tetrahedron, also provides three non-coplanar surfaces $S_{IJ}$, bounded
by the edges $e_{I}$, $e_{J}$ and the `opposite' edge. We have thus a
natural choice of surfaces and edges. For these surfaces and edges, we
have, $\Phi_{IJ} \approx \Case{1}{2}\mathbb{P}^a{\cal
E}_{abc}(\dot{e}^b_I\delta t_I)(\dot{e}^c_J\delta t_J)$ and
$h_{K}^{-1}\{h_K, \sqrt{{\cal V}_n}\} \approx \{\delta t_K \dot{e}^a_K
A_a^i\tau_i, \sqrt{{\cal V}_n}\}$.  Clearly
\[
\frac{1}{6}{\cal E}^{IJK} \Phi_{IJ} h_{K}^{-1}\big\{h_K, \sqrt{{\cal V}_n}\big\}
 \approx  \frac{1}{12}{\cal E}^{IJK}\mathbb{P}^a{\cal E}_{abc}(\delta
t_I\dot{e}_I^b)(\delta t_J\dot{e}_J^c)(\delta t_K \dot{e}^d_K)
\big\{A^i_d\tau_i , \sqrt{{\cal V}_n}\big\}.
\]
Now
\[
{\cal E}^{IJK}\dot{e}^b_I\dot{e}^c_J\dot{e}^d_K = {\cal
E}^{bcd}\det(\dot{e}^a_I),\qquad \det (\dot{e}^a_I)\delta
t_1\delta t_2\delta t_3 =: \epsilon^3 ,\qquad  {\cal E}_{abc}{\cal
E}^{bcd} = 2\delta^d_a.
\]
Using these, it follows
\begin{gather*}
\frac{1}{8\pi}\int_{\Sigma_3} N
\frac{q_{ab}}{\sqrt{q}}\mathbb{P}^a\mathbb{P}^b  \approx
\frac{2}{\kappa^2\gamma^2\pi}\sum_n N(v_n)\sum_{\Delta_n} \delta_{ij}
\big(\epsilon_n^3\mathbb{P}^c\delta^a_c \big\{A_a^i, \sqrt{{\cal
V}_{\Delta}}\big\}\big) \big(\epsilon_n^3\mathbb{P}^d\ \delta^b_d\
\big\{A_b^j, \sqrt{{\cal V}_{\Delta}}\big\}\big) \nonumber \\
\hphantom{\frac{1}{8\pi}\int_{\Sigma_3} N
\frac{q_{ab}}{\sqrt{q}}\mathbb{P}^a\mathbb{P}^b }{}
 =   \left(- \frac{4}{\kappa^2\gamma^2\pi}\right)
\sum_{n} N(v_n) \sum_{\Delta_n}\frac{{\cal E}^{IJK}}{6}\frac{{\cal
E}^{I'J'K'}}{6} \mathbb{\Phi}_{IJ} \mathbb{\Phi}_{I'J'} \nonumber \\
\hphantom{\frac{1}{8\pi}\int_{\Sigma_3} N
\frac{q_{ab}}{\sqrt{q}}\mathbb{P}^a\mathbb{P}^b =}{}
 \times\mathrm{Tr}\left[
h_{K}^{-1} \big\{ h_{K}, \sqrt{{\cal V}_{\Delta}}\big\}
h_{K'}^{-1} \big\{ h_{K'}, \sqrt{{\cal V}_{\Delta}}\big\} \right].
\end{gather*}

In the last equation, we have included $\mathrm{Tr}\,\tau_i\tau_j = -
\Case{1}{2}\delta_{ij}$. All the factors of $\epsilon$'s have been
neatly absorbed, again thanks to density weight 1 of the Hamiltonian.
The sum over $n$ is a sum over vertices of the triangulation. The sum
over $\Delta_n$ includes those tetrahedra whose distinguished vertex is
$v_n$ and this is a f\/inite sum. Note that some vertices of the
triangulation may not be distinguished vertices of {\em any}
tetrahedron.  Such vertices do not contribute to the sum. Therefore, the
sums over $v_n$ and $\Delta_n$ together, denote sum over the cells of the
decomposition.  Notice that surfaces over which the {\em electric flux}
is taken is correlated with the edge along which the {\em gravitational
holonomy} is to be taken thanks to the ${\cal E}^{IJK}$ factors.

In the passage to quantum theory, the triangulation is adapted to the
underlying graph of a cylindrical function\footnote{The cylindrical
functions in this case are analogous to the gravitational spin network
functions except for the replacement of the spin label by a  {\em
integer charge} label with charge conservation at each vertex and are
called {\em charge network} functions \cite{Madhavan}. These have been
called {\em flux networks} in \cite{ALReview}.} by aligning the edges of
the elementary tetrahedra to be segments of corresponding edge of the
graph and the distinguished vertex to be the vertex of the graph
\cite{ThiemannScalar,ThiemannQSD,ThiemannQSD2,ThiemannQSD4}. Upon quantization, only those tetrahedra will
contribute which have an edge overlapping with the edge of the holonomy.
In each contributing cell (tetrahedron), the action of the electric f\/lux
operator, $\mathbb{\Phi}_{IJ}$ will be non-zero {\em only} on the
Maxwell holonomy along the edge~$e_{K}$ which is transversal to the
surface $S_{IJ}$ and will give $\hbar q_K$. This also selects the edge
$K$ in the gravitational holonomy, $h_{K}$. Here $q_K$, an integer, is
the charge (in units of electric charge) in the Maxwell holonomy. On any
cylindrical function, the action of the regulated quantum operator,
results in a f\/inite sum of terms since only vertices of the underlying
graph contribute to the sum, regardless of ref\/inement of the
triangulation.

We have deviated somewhat from the procedure given in \cite{ALReview,ThiemannScalar,ThiemannQSD,ThiemannQSD2,ThiemannQSD4}. We have not done the point splitting and explicitly made
passage to the f\/lux operators absorbing the factors of $\epsilon$'s (in
the limit of small $\epsilon$). The correlation between the edge in the
gravitational holonomy and the surface in the Maxwell f\/lux is preserved
through the use of triplet of edges and corresponding unique transversal
surfaces provided by an elementary tetrahedron.

For the magnetic term, we can proceed identically by def\/ining a f\/lux
$\mathbb{\Psi}_S := \int_S \mathbb{B}^a{\cal E}_{abc}dS^{bc}$ which will
lead to an equation same as the one above with $\mathbb{\Phi}_S$
replaced by $\mathbb{\Psi}_S$. Unlike the electric f\/lux however, this
magnetic f\/lux is {\em not} an elementary variable. For
\begin{gather*}
\mathbb{\Psi}_S   :=   \int_S \frac{1}{2}{\cal
E}_{abc}\mathbb{B}^adx^b\wedge dx^c
 =   \int_S {\cal E}_{abc}\frac{{\cal
 E}^{aef}}{4}\mathbb{F}_{ef}dx^b\wedge dx^c \nonumber \\
\phantom{\mathbb{\Psi}_S }{}  =  \int_S (\delta^e_b \delta^f_c - \delta^e_c \delta^f_c)\frac{1}{4}
\mathbb{F}_{ef} dx^b\wedge dx^c
=   \int_S \mathbb{F}   \approx  \frac{1}{2}\left(\mathbb{h}(e_S) -
\mathbb{h}^{-1}(e_S)\right),
\end{gather*}
where $e_S$ is a small close curve bounding the surface $S$.  When the
triangulation is adapted to the underlying graph of a cylindrical
function, the curve~$e_S$ may be taken to be one of the loops~$\alpha_{IJ}$ of Thiemann \cite{ThiemannScalar,ThiemannQSD,ThiemannQSD2,ThiemannQSD4}, which bounds the face~$S_{IJ}$. Because of the presence of the Maxwell holonomies (with unit
charge), the underlying graph of a cylindrical state is changed by
adding these loops.  But there are no gravitational holonomies along
these loops and hence the spin labels of these extra loops are zero.

We note in passing that if we have a spin and charge network state, i.e.\
a graph with both the spin and the charge labels for its edges some of
which could be zero, then the action of the electric part of the
Hamiltonian will be {\em non-zero only on those edges which have both
the spin and the charge labels to be non-zero}~-- the volume operator
will kill edges with no spin while the electric f\/lux operator will kill
the edge with no charge. This feature enables the operator to be well
def\/ined on the kinematical Hilbert space even in the limit of inf\/inite
ref\/inement. By contrast, the magnetic part will {\em not} kill an edge
with zero charge but will add loops with unit charge. This prevents the
operator to be well def\/ined on the kinematical Hilbert space in the
limit of inf\/inite ref\/inement.  It could however be def\/ined on a subspace
of the algebraic dual of the space of the cylindrical functions~\cite{ALReview}.

\subsection[Yang-Mills field theory]{Yang--Mills f\/ield theory}
Apart from the non-Abelian nature of the gauge f\/ield, in form, the
expressions proceed in the same manner as for the Maxwell case. The
details are given in \cite{ThiemannScalar,ThiemannQSD,ThiemannQSD2,ThiemannQSD4}.

\section{Propagation of matter waves on quantum geometry}
\label{MatterWaves}

While there are good arguments to justify search for a quantum gravity,
its observable signatures have been quite dif\/f\/icult to come by. Various
dif\/ferent approaches to quantum gravity have contained a suggestion that
space-time of quantum gravity is dif\/ferent from the continuum classical
geometry be it discreteness and/or non-commutativity etc. Coupled with a
heuristic idea that existence of a fundamental length scale may indicate
a conf\/lict with Lorentz invariance, a possible violation of Lorentz
invariance has been considered as a possible signature of quantum
gravity. The violation is thought to be manifested as a deviation from
the Lorentz invariant {\em dispersion relation} for particle propagation
on a `quantum space-time'. Its cumulative ef\/fect over cosmological
distances are estimated to be detectable. From LQG Hamiltonian
constraint, such deviation have been extracted \cite{Alfaro,Morales,Morales2,Pullin} using the following set of assumptions\footnote{The subject of
Lorentz invariance violation in all its aspects is rather large. We have
included it here only to the extent of indicating how a computation of
potentially observable signatures involving matter could be done. For
review, we refer the reader to~\cite{LorentzViolation}.}. The Lorentz
violations in the Polymer quantized scalar in Minkowski background has
already been discussed in Section~\ref{PolymerScalar}.

One begins by assuming a quantum state to be of a product form with a
suitable state for geometry as well as for matter. For example, the
early works, e.g.~\cite{Pullin}, the geometry state was a {\em weave}
state~\cite{Weaves,Weaves3,Weaves6,Weaves7,Weaves4,Weaves5,Weaves2} which provides a length scale $L_{\rm weave} \gg \lP$
such that for larger scales the 3-geometry may be approximated by a
continuum while the discrete structure manifests near the Planck scale.
The scales relevant for observations of interest are assumed to be much
larger than~$L_{\rm weave}$. The matter state is chosen to be a coherent
state so that the expectation values of the matter operators can be
approximated by classical f\/ields (f\/ields obeying classical equations to
leading order). These f\/ields are assumed to be varying slowly over a
length scale of~$\lambda \gg L_{\rm weave}$. Taking expectation value of the
Hamiltonian constraint, the geometry operators are replaced by the
metric used in the construction of the weave plus corrections of the
order of $\lP/L_{\rm weave}$. The expectation values of the matter part of
the Hamiltonian constraint is taken as def\/ining an {\em effective
Hamiltonian}. The {\em space-time metric} is taken to be a static metric
determined by the 3-metric and a lapse~$N$. The matter equations of
motion following from the ef\/fective Hamiltonian contain the
modif\/ications implied by quantum gravity. In computing the ef\/fective
Hamiltonian, one expands the classical f\/ields at the vertices in a
region of size~$L_{\rm weave}$ around a central point of the region and
averages over these vertices. The symmetries of the 3-geometry (eg
rotational invariance of the f\/lat geometry) is used to restrict the form
of the averages. In turns out that there are corrections in the
quadratic order in the Maxwell f\/ields which imply polarization dependent
speed of propagation suggesting birefringence. The computations are
varied by dif\/ferent choice of states for expectation values eg more
general semi-classical states rather than weaves or even using only
qualitative properties of `would be semi-classical states'~\cite{Alfaro}.  These computations also give additional corrections.
Subsequently, the method is also applied to propagation of neutrinos~\cite{Morales,Morales2}.  These early computations show how Lorentz invariance
violations~-- as manifested by dispersion relations~-- could potentially
arise in LQG. But do they {\em imply} that LQG violates Lorentz
invariance?

This is not easy to answer. The above computations are essentially at a
kinematical level. This however need not be regarded as a strong
drawback, because the very characterization of Lorentz violation
pre-supposes a background space-time in which one can f\/ind locally
inertial observers. Introducing a background is akin to a gauge f\/ixing
and heuristically at least the use of kinematical states could be
justif\/ied.  There are still not satisfactory semi-classical states of
the geometry and matter system which will describe `test f\/ields' on a
given space-time background. The proper framework for this would be
quantum f\/ield theory on curved space-times. To obtain it from the full
quantum theory of gravity and matter, is a challenging task. First steps
in this direction are discussed by Sahlmann and Thiemann~\cite{ThiemannSahlmann,ThiemannSahlmann2}.  A cautionary note on a dif\/ferent aspect, has
also been sounded in~\cite{BojowalMoralesSahlmann}, pointing out that
staying within Hamiltonian framework for computation including {\em
approximation} leading to higher order derivatives, could be misleading
because the corrections could be due to additional degrees of freedom
with a Lorentz invariant dynamics. There is yet another aspect.
Dispersion relation, relativistic or otherwise, need a~background
space-time. The presumed space-times in the f\/irst approaches have been
static space-times with spatial geometry being described by a
semi-classical state. Thus the `time' part of the geometry does not have
anything to do with the quantised spatial geometry. One has not used a
quantum space-time, only quantum space. An extension to quantum
space-time has been considered in the context of FLRW cosmology~\cite{QFTonQG}. Such extensions are probably needed to discuss matter
wave propagations on quantum space-time and its implications.

\section{Concluding remarks}\label{Summary}

We began by noting that non-Fock quantization of matter f\/ields is
necessary once `space-time' is dynamical. However, the illusion of a
f\/ixed background space-time with Fock quantized f\/ields on it, is also
extremely persuasive. In these circumstances, gravitational interactions
of matter f\/ields are negligible compared to intra-matter interactions.
Consequently, the observed semi-classicality of geometry itself would
conceivably be sensitive to matter content as well as their
interactions. Matter interactions must play a role in getting our `late
universe' with its essentially classical space-time emerge from the
quantum gravity regime of the early universe. Study of (loop) quantized
matter has the potential to provide selection criteria from the
stability of semi-classical states of geometry. We are still quite far
away from this goal.

There are other `applications' of matter f\/ields. They can serve as
probes for understanding properties of background independent
quantization. For instance, in a f\/ixed background space-time such as the
Minkowski space-time, the short distance behaviour of quantum f\/ields
play a~role in admitting the chiral (and the trace) anomalies. One of
these, namely the axial anomaly, has played a~historic role in
understanding the $\pi^0 \to 2 \gamma$ process. Cancellation of gauge
anomalies has also provided constraints on model building. In a
background independent quantization, we loose the handle of short
distance behaviour. How then are the anomalies to be understood? This is
an old issue more recently mentioned by Nicolai et al.~\cite{Nicolai},
still awaiting an understanding.

The focus of analysis of matter in quantum gravity has been to study its
propagation in the regime of semi-classical gravity. These have
indicated that one could expect dispersion relations not compliant with
Lorentz invariance. This has been seen in the propagation of massless
fermions (neutrinos), light as well as massless scalar f\/ields. For
fermions and light, quantum ef\/fects of spatial geometry are incorporated
while space-time is constructed classically. The case of the scalar
f\/ield discussed in Section~\ref{PolymerScalar}, is qualitatively
dif\/ferent from these. Here the relativistic (massless) scalar f\/ield is
classically expressed as a collection of classical `harmonic
oscillators' with frequencies $\omega_{{\bf k}} \sim |{\bf k}|$ and
there is no Lorentz violation at this stage. For the usual Fock
quantization, the energy spectrum, in units of $\hbar\omega_{{\bf k}}$
is linearly spaced and the Lorentz invariant dispersion results from the
propagator. With polymer quantization, the spectrum is not linearly
spaced and leads to Lorentz violating dispersion relation. While one can
construct Lorentz generators from the $\pi_{{\bf k}}$, $\phi_{\bf k}$
basic variables, one cannot do so with the $U_{\lambda{\bf k}} -
U^{\dagger}_{\lambda{\bf k}}$ variables for $\lambda \neq 0$. Clearly,
it is the polymer quantization which requires the use of $U_{\lambda{\bf
k}}$ with non-zero $\lambda$ which is the source of this Lorentz
violation at the level of `free f\/ield'. Importing the Lorentz violating
dispersion relation into the usual computational scheme of perturbative
quantum f\/ield theory based on {\em Fock} quantization with its UV
divergences (especially the `quadratic divergences'), will lead to the
`f\/ine tuning' problem discussed by Collins et al.~\cite{CollinsPerezSudarsky} (see also~\cite{JainRalston}). To check the
viability of a polymer quantized f\/ield theory on a Minkowski background,
one will need to develop a computational scheme for a polymer quantized,
interacting theory and then test if at `low energies' Lorentz violating
ef\/fects are suppressed or not.  Currently this lies only in the realm of
possibilities.

\subsection*{Acknowledgements}

G.D.\ would like to thank for Romesh Kaul for a
discussion on the f\/ine tuning problem.

\pdfbookmark[1]{References}{ref}
 \LastPageEnding

\end{document}